\documentclass[
 pra, twocolumn,
 superscriptaddress,
 showpacs,
 longbibliography,
 aps,
 showkeys]{revtex4-1}

\usepackage[pdftex]{hyperref,color,graphicx}
\usepackage{amsfonts,amssymb,amsmath}
\usepackage[separate-uncertainty=false]{siunitx}
\usepackage[english]{babel}
\usepackage[utf8]{inputenc}
\usepackage[T1]{fontenc}

\usepackage{graphicx}
\usepackage[space]{grffile}
\usepackage{latexsym}
\usepackage{textcomp}
\usepackage{longtable}
\usepackage{multirow,booktabs}
\usepackage{url}
\hypersetup{colorlinks=false,pdfborder={0 0 0}}
\usepackage[english]{babel}
\usepackage{dcolumn}
\usepackage{bm}
\usepackage{units}
\usepackage{upgreek}
\usepackage{color}
\usepackage{cancel}
\usepackage{hyperref}
\usepackage{braket}

\makeatletter
\renewcommand{\p@subsection}{}
\renewcommand{\p@subsubsection}{}
\makeatother

\begin{document}

\title{ Experimental realization of a momentum-space quantum walk }

\author{Siamak Dadras}
\affiliation{Department of Physics, Oklahoma State University, Stillwater, Oklahoma 74078-3072, USA}

\author{Alexander Gresch}
\affiliation{Faculty of Mathematics and Natural Sciences, Heinrich Heine Universit\"at D\"usseldorf, 40225 D\"usseldorf, Germany}

\author{Caspar Groiseau}
\affiliation{Dodd-Walls Centre for Photonic and Quantum Technologies, Department of Physics, University of Auckland, Private Bag 92019, Auckland, New Zealand}

\author{Sandro Wimberger}
\email{sandromarcel.wimberger@unipr.it}
\affiliation{Dipartimento di Scienze Matematiche, Fisiche e Informatiche, Universit\`{a} di Parma, Parco Area delle Scienze 7/A, 43124 Parma, Italy}
\affiliation{INFN, Sezione di Milano Bicocca, Gruppo Collegato di Parma, Parma, Italy}

\author{Gil S. Summy}
\email{gil.summy@okstate.edu}
\affiliation{Department of Physics, Oklahoma State University, Stillwater, Oklahoma 74078-3072, USA}

\begin{abstract}
We report on a discrete-time quantum walk that uses the momentum of ultra-cold rubidium-87 atoms as the walk space and two internal atomic states as the coin degree of freedom. Each step of the walk consists of a coin toss (a microwave pulse) followed by a unitary shift operator (a resonant ratchet pulse). We carry out a comprehensive experimental study on the effects of various parameters, including the strength of the shift operation, coin parameters, noise, and initialization of the system on the behavior of the walk. The walk dynamics can be well controlled in our experiment; potential applications include atom interferometry and engineering asymmetric walks.  
\end{abstract}




\maketitle

\section{INTRODUCTION}\label{sec:intro}

The idea that quantum computational devices can be more powerful than their classical versions has been the incentive for research in the last three decades to adopt classical algorithms in quantum computation \cite{feynman1982simulating,nielsen2010quantum,shor1994algorithms,travaglione2002implementing}. In this respect, the quantum walk (QW) has been a workhorse in devising new quantum algorithms that are more effective than their classical counterparts \cite{DT-QC}. This includes, for instance, an algorithm which can search an unsorted database quadratically faster than any classical version \cite{grover1997quantum}, or a quantum walk on a hypercube with exponentially faster hitting time as compared to a classical random walk \cite{yamasaki2002analysis,kempe2005discrete,QWBook}. 


QWs are, in principle, the extension of the idea of classical random walks in quantum mechanics, describing the propagation of quantum particles on periodic potentials \cite{aharonov1993quantum,farhi1998quantum,kempe2003quantum}. The QW we present here is an adoption of the discrete-time classical walk into the quantum regime, and hence it consists of two degrees of freedom: a space in which the walk takes place, and a ``coin'' which selects the path of the system through the walker's space \cite{aharonov1993quantum, ambainis2001one, kempe2003quantum}. While the basic procedure for producing a QW is outwardly analogous to its classical counterpart, the propagation dynamics are totally different. 

\begin{figure}
	\begin{center}
		\includegraphics[width=\linewidth]{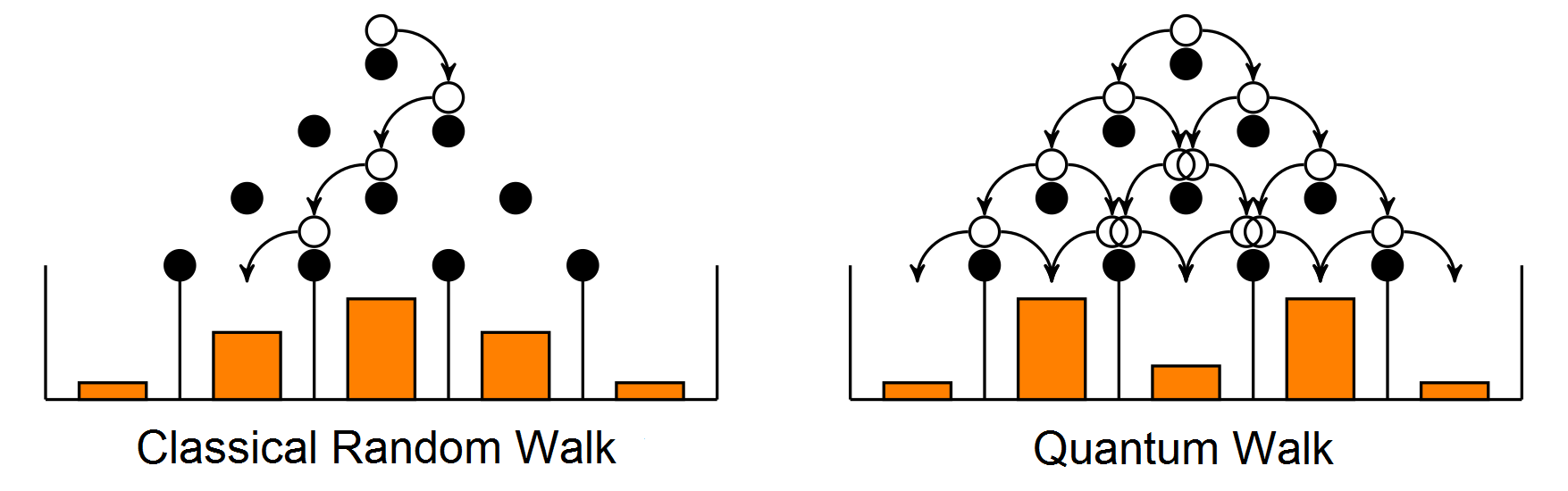}
		\caption{Sketch of the difference between a classical Galton board walk (left panel) and its quantum version (right panel). In the latter all paths interfere to give the final distribution at the end of the walk that typically shows two characteristic peaks moving ballistically with the number  from the center \cite{kempe2003quantum}.} \label{fig:saw0}
	\end{center}
\end{figure}

Unlike in classical walks, a quantum walker can be in a superposition of two (or more) states that can take all possible paths through the walk space simultaneously, see Fig.~\ref{fig:saw0}. Thus, there exists an interference between the multitude of paths and an entanglement between the two degrees of freedom in a QW \cite{dadras2018quantum, summy2016quantum}, leading to faster propagation and enhanced sensitivity to initial conditions \cite{preiss2015strongly}. These features have raised considerable interest in using QWs as building blocks in probabilistic algorithms for universal quantum computing \cite{shenvi2003quantum, childs2009universal, childs2013universal,
childs2003exponential, childs2004spatial, ambainis2005proceedings, DT-QC} and for quantum information processing \cite{venegas2012quantum, QWBook}. 

In contrast to several other QWs experimentally implemented in an assortment of walk spaces with a variety of possible walkers species \cite{preiss2015strongly,dur2002quantum,eckert2005one,
steffen2012digital,groh2016robustness,karski2009quantum,
chandrashekar2006implementing,travaglione2002implementing,
zahringer2010realization,schmitz2009quantum,
wineland1998experimental,perets2008realization,
peruzzo2010quantum,cardano2017detection,
chen2018observation,wang2018experimental}, we realize our walks in momentum space with a spinor Bose-Einstein condensate (BEC) of $^{87}$Rb atoms. We demonstrate one of the characteristic signatures of a QW, with peaks in the walk distribution which propagate ballistically away from the origin, leading to a standard deviation of the walk distribution growing proportionally to the number of steps as the walk proceeds. This feature proves the quantum nature of the walk as compared to a classical diffusive walk \cite{kempe2003quantum, QWBook}. We conduct several experiments to investigate the effects of various parameters on the dynamics of our discrete-time QW. This includes looking at the effects of the kick strength responsible for the shift operation, coin choice, phase adjustment, and the BEC characteristics on the behavior of this walk. By introducing an engineered noise to the system, we also show how the classical effects start to emerge and dominate the walk toward the expected classical distribution.         

An advantage of implementing a QW in momentum space is that one can have a robust control on both internal and external degrees of freedom of the walker. This stability arises naturally from the creation and control of a BEC, which is typically done in momentum space \cite{Pethick, Stringari}.
In our system these degrees of freedom are atomic hyperfine states and the center-of-mass momentum of the atoms in a pulsed optical lattice. By manipulating either degree of freedom we show how the walk can be steered or even reversed as desired. 
 
The paper is organized as follows: The next two sections briefly review the theory of the discrete-time QWs (Sec. \ref{sec:QW}) and details on their experimental implementation (Sec. \ref{sec:exp.}). Section \ref{sec:results} presents our results, with all the aspects of control on the QW given in the corresponding subsections. Section \ref{sec:concl.} concludes the paper.
 
 \section{DISCRETE-TIME QUANTUM WALKS IN MOMENTUM SPACE}
\label{sec:QW}

Each step of our QW, ${\hat{\bf{U}}}_\mathrm{step}={\hat{\bf{T}}}{\hat{\bf{M}}}$, consists of a coin operator ${\hat{\bf{M}}}$ which produces a superposition of two internal states, followed by a unitary shift operator ${\hat{\bf{T}}}$, which entangles the internal and external degrees of freedom. We realize the coin operator using resonant microwave (MW) radiation that addresses the walker's internal degree of freedom, i.e., the two ground hyperfine levels of the $^{87}$Rb atoms. These two levels are the $F=1,~m_F=0$, and the $F=2,~m_F=0$, states, hereafter denoted by $|1\rangle$ and $|2\rangle$ respectively. 
The coin operator is represented by the unitary rotation matrix
\begin{equation}
{\hat{\bf{M}}}({\alpha},{\chi})=
\begin{bmatrix}
& \cos{({\alpha}/2)}& {e^{-i\chi}}{\sin({\alpha}/2)}\\
& {-e^{i\chi}}{\sin({\alpha}/2)} & \cos{({\alpha}/2)}\\
\end{bmatrix},
 \label{eq1} 
\end{equation}
in which $\alpha$ and $\chi$ are the polar and azimuthal precessions on the Bloch sphere. A 50:50 coin operation corresponds to a $\pi/2$ ($\alpha=\pi/2$) pulse of the MWs. 
To make the step direction of our walk contingent upon the result of each coin toss, we apply the unitary shift operator, 
\begin{equation}
\hat{\bf{T}} = \exp{(iq{\hat{\theta}})} |1\rangle \langle 1| + \exp{(-iq{\hat{\theta}})} |2\rangle \langle 2|,
 \label{eq2} 
\end{equation}
which shifts the momentum by $\pm{q}$ depending on whether the atom resides in the internal state $|1\rangle$ or $|2\rangle$. This produces a strong mixing between internal and external degrees of freedom at each step of the walk. In Eq. (\ref{eq2}), $q$ is an integer in units of two-photon recoils ${\hbar}G$ ($G=2\pi/\lambda_G$ with $\lambda_G$ being the spatial period of the standing wave implementing the walk space in our setup), and ${\hat{\theta}}=\hat{x}\mod(2\pi)$ where $\hat{x}$ is the dimensionless position operator. In a standard walk, $q=1$, corresponding to nearest-neighbor coupling in momentum space. This gives the ideal shift operator in matrix representation as
\begin{equation}
\hat{\bf{T}}=
\begin{bmatrix}
|n+1\rangle \langle n| &0\\
0 & |n-1\rangle \langle n|
\end{bmatrix},
\label{eq3}
\end{equation}
which can shift the external (momentum) eigenstates $\{|n\rangle\}, n\in \mathbb{Z}$ by +1 or -1 unit when applied on either $|1\rangle$ or $|2\rangle$ internal state.  

Our shift operator is a quantum resonant ratchet \cite{Lundh2005, sadgrove2007, gil2008} based on the atom-optics kicked rotor (AOKR), which has been used extensively in quantum chaos experiments (see, for example, \cite{raizen1999quantum, ni2017hamiltonian, SW2011}). AOKR experiments work with ultra-cold atoms subject to a series of short pulses of a 1D off-resonant optical lattice (standing wave). In contrast to previous work, here we adjust the frequency of our lattice laser to be halfway between the $|1\rangle$ and $|2\rangle$ levels at the ground state to the $5^2P_{3/2}, F = 3$ level at the excited state. Thus these internal states see a lattice detuning of equal size but with opposite signs. Using dimensionless variables, the dynamics of the AOKR are described by the Hamiltonian \cite{raizen1999quantum}, 
\begin{equation}
\label{eq4}
{\hat{H}}({\hat{x}},{\hat{p}}_{x},t)={{\hat{p}}_{x}}^2/2+k\left[1+\cos({\hat{x}})\right]\sum_{j{\in}{\mathbb{Z}}}{\delta}(t-j\tau),
\end{equation}
where $j$ counts the number of pulses, $t$ and $\tau$ are the dimensionless time and pulse period, and ${\hat{p}}_{x}$ is the rescaled momentum operator.  The pulse strength is $k={\Omega}^2{\tau_p}/\Delta$, in which ${\tau_p}({\ll}\tau)$ is the pulse length, $\Omega$ is the Rabi frequency, and $\Delta$ is the detuning of the laser light from the atomic transition. The factor $1+\cos({\hat{x}})$ represents the spatial periodicity of the potential due to the interference of the lattice laser beams. Using Bloch theory \cite{kittel1996introduction}, the dynamics of an AOKR is described in the Raman-Nath regime \cite{gould1986diffraction} by a one-period evolution operator called the Floquet operator \cite{SW2011}
\begin{equation}
\label{eq:saw1}
{\hat{\mathcal{U}}}_{{\hat{p}_{x}},k}={\hat{\mathcal{U}}}_f{\hat{\mathcal{U}}}_k
=e^{-i{\tau}{{\hat{p}}_{x}}^2/2}e^{-ik\left[1+\cos({\hat{\theta}})\right]} \,. 
\end{equation}
Here, ${\hat{\mathcal{U}}}_f$ signifies the free evolution between two pulses, and 
\begin{equation}
\label{eq:saw2}
{\hat{\mathcal{U}}}_k=e^{-ik}\sum_{m}(-i)^{m}{\exp}(-im{\hat\theta)}J_{m}(k)\,, 
\end{equation}
with $J$'s being Bessel functions of the first kind, represents the ``kick'' by each pulse that leads to a symmetric diffraction of the wave function in the space spanned by the momentum eigenstates \cite{Weiss2015}. The prefactor $e^{-ik}$, which originates from the dc component of the lattice potential, introduces a ``global'' phase to the atoms at each kick \cite{dadras2018discrete, caspar2018}. As will be shown later, we can compensate for this phase in our experiments. 

\begin{figure}
	\begin{center}
		\includegraphics[width=\linewidth]{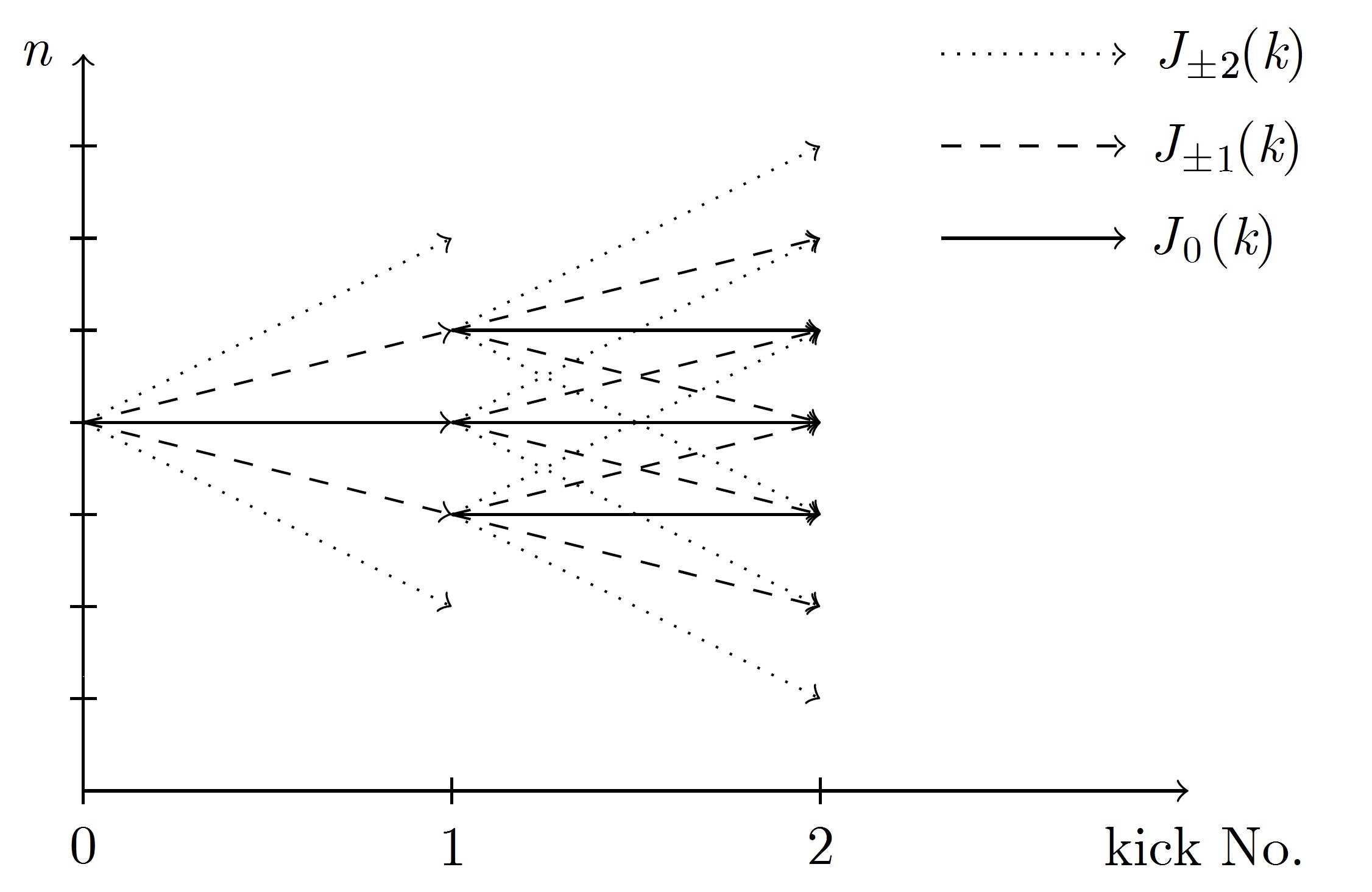}
		\caption{Sketch of diffraction into various momentum classes by the application of two successive optical lattice kicks from an initial state consisting of a single momentum state. The various lines represent the different couplings produced when $k$ is changed.} \label{fig:saw1}
	\end{center}
\end{figure}

The formula from Eq.~\eqref{eq:saw2} shows that in our walk various momentum classes are, in principle, coupled to each other in one step or kick. By carefully choosing the argument of the Bessel function (the kick strength), these couplings can be controlled to some extent. The result is an effective nearest-neighbor coupling in the momentum classes, or a diffraction mainly into the first-order peaks when the kick strength is approximately 1. From previous studies \cite{CasparMA, summy2016quantum} and our experimental results to be shown below we understand that the values for realizing a walk with ballistically moving peaks, and hence a standard deviation proportional to the number of steps, lie in the range $k \in [1.4, 2]$. The optimal value for maximized nearest-neighbor coupling and most similarity with a standard quantum walk is $k \approx 1.5$, as then the first-order diffraction starts to dominate the zeroth order, while the second order still remains relatively weak. A sketch of the diffraction into the different momentum classes by successive applications of the kick is found in Fig.~\ref{fig:saw1}.

While the AOKR itself can also produce ballistic expansion, the quantum resonant motion in its case would be symmetric and equal for both internal states \cite{Izr1990, SW2011, Weiss2015}. We implement one of the requirements of a discrete-time QW by a quantum resonance ratchet effect \cite{Lundh2005, sadgrove2007, gil2008}, such that the two internal states move in opposite directions during the kick evolution. The use of two internal degrees of freedom in the AOKR evolution was previously proposed in Refs. \cite{Scharf1989, Michael2003} and also in the context of QWs \cite{Simon2006, R2013}. However,  as will be demonstrated in Sec. \ref{sec:results}, our implementation gives exceptional experimental control over the initial conditions and system parameters. To realize the simplest quantum resonant ratchet, we should meet the quantum resonance condition so that the effect of the free evolution due to the kinetic energy of the atoms (i.e., ${\hat{\mathcal{U}}}_f$ in the Floquet operator) \cite{SW2011} is eliminated. This can be seen in the AOKR as the Talbot effect (in the time domain) for atomic matter waves diffracted from a phase grating induced by a pulsed optical lattice \cite{SW2011}. We fulfill this condition by choosing $\tau=4\pi$ (Talbot time) for our dimensionless pulse period, which is also experimentally long enough to allow the delivery of the coin-toss MW radiation between the ratchet pulses. 

The quantum resonant ratchet effect requires breaking the spatial-temporal symmetry of the problem \cite{SW2011, white2013experimental, ni2017hamiltonian, Lille2018} to direct the dynamics of the AOKR in momentum space accordingly. Experimentally, this is achieved by the choice of an initial state $1/\sqrt{2}(|n=0\rangle+e^{i\phi}|n=1\rangle)$ created by a long pulse of the off-resonant standing wave (Bragg pulse) on the original BEC ($|n=0\rangle$) \cite{ni2016initial,ni2017hamiltonian}. Applying the AOKR to this state, results in a change in the average momentum by an amount ${\Delta}\langle{\hat{p}}\rangle=-k\sin{(\phi)}/2$ after each pulse \cite{SW2011}. The key point to understanding our walk is that, since $k\propto{1/\Delta}$, the two internal states undergo ratchets in opposite directions (the detuning has opposite signs for each state). Thus by choosing $\phi=\pi/2$ and $|k|{\sim}1.5$, one can increase or decrease (depending on the sign of $k$) the average momentum of either internal state at each step of the ratchet by one unit. 

\section{EXPERIMENTAL PROCEDURE}
\label{sec:exp.}

\begin{figure}
	\begin{center}
		\centering
		\includegraphics[scale=0.32]{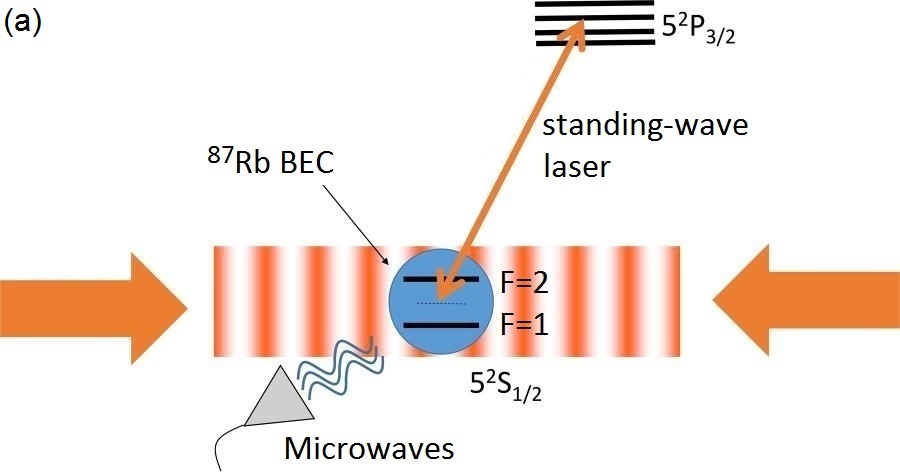}
		\vspace{0.3cm}
		\includegraphics[scale=0.68]{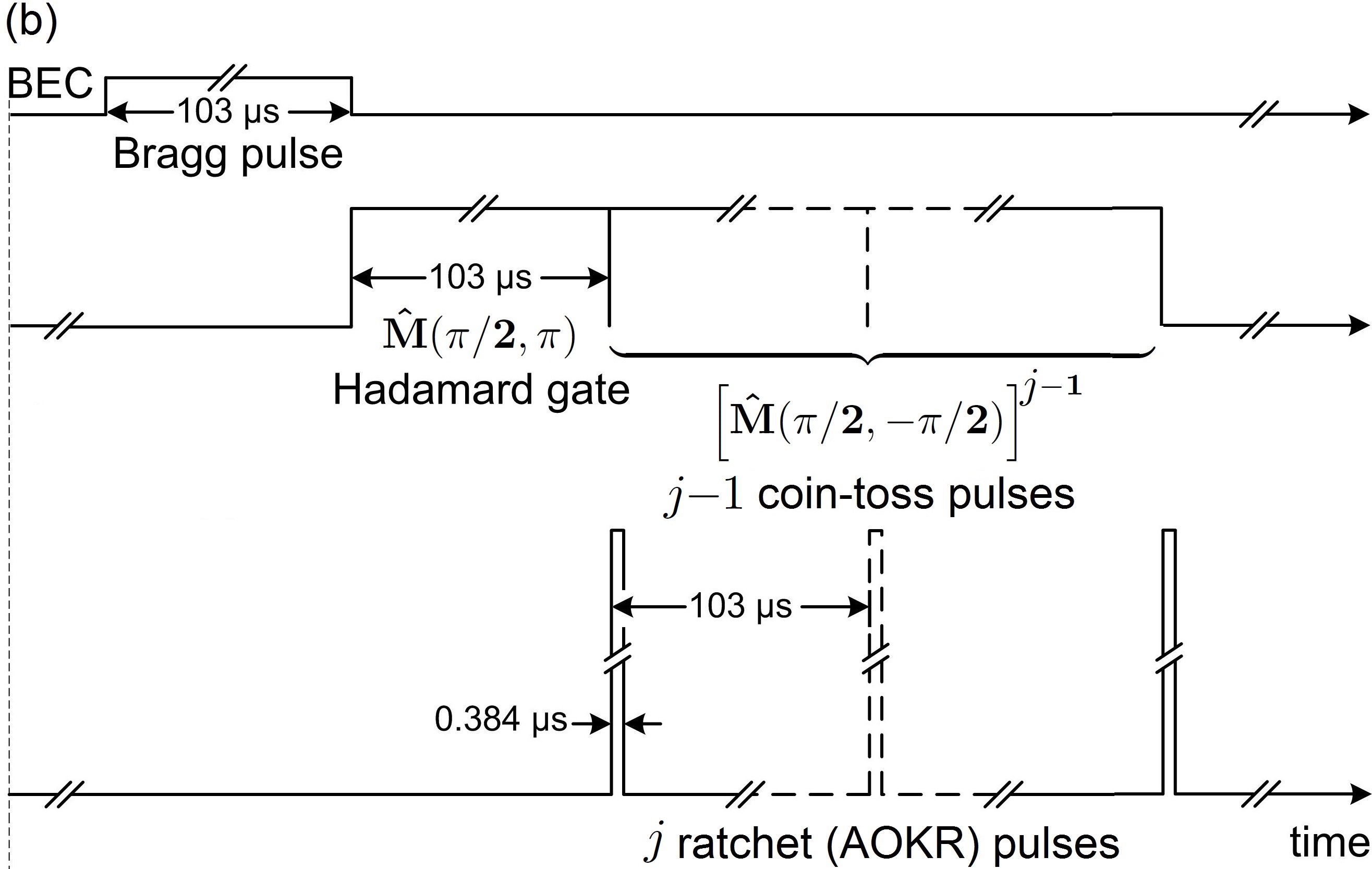}
\caption{(a) Schematic of our experiment for the realization of a QW in momentum space, adapted from \cite{summy2016quantum}. The optical lattice is applied periodically in a quantum resonance ratchet scheme to implement the momentum shifts. The internal states $|1\rangle$ and $|2\rangle$ of the $^{87}$Rb BEC atoms are addressed by $\sim$6.8 GHz MW pulses out of a C-band horn antenna. The standing wave was created by two laser beams with a wavelength of $\sim$ 780~nm, $\sim$ 3.4~GHz red detuned from the $5^2S_{1/2}, F=1\rightarrow 5^2S_{3/2}, F'=3$ D$_2$ transition, to address the internal states with equal $|k|$. (b) Timing scheme of the Bragg, ratchet, and MW pulses in our QW experiments. Duration of the Bragg pulse and the time between ratchet pulses are adjusted to be 103$\mu$s. This (Talbot) time satisfies the quantum resonance ratchet conditions in our experiments and is also sufficiently long to allow for the application of $\pi/2$ MW pulses.} \label{fig1}
	\end{center}
\end{figure}

To conduct the QW experiments, a BEC of about 70,000 $^{87}$Rb atoms in the $5^2S_{1/2}, F=1, m_F=0$ state is created by evaporative cooling of optically trapped atoms in a focused CO$_2$ laser beam. This procedure, including details of our magneto-optical trap, optical molasses, and evaporative cooling is described elsewhere \cite{barrett2001all}. A schematic of our experimental setup is shown in Fig.~\ref{fig1}(a). Immediately after the BEC is released from the dipole trap, the QW sequence is applied on the atoms as shown in Fig.~\ref{fig1}(b).
To carry out a standard walk with $j$ steps, we apply the sequence $(\hat{\bf{U}}_{step})^j=[{\hat{\bf{T}}}{\hat{\bf{M}}}(\pi/2,-\pi/2)]^{j-1}[{\hat{\bf{T}}}{\hat{\bf{M}}}(\pi/2,\pi)]$ to the initial state $|\psi_0\rangle=|1\rangle\otimes1/\sqrt{2}(|n=0\rangle+ i |n=1\rangle)$ prepared by a Bragg pulse. Note that in the first step, we use a different coin toss, the so-called Hadamard gate, to prepare the internal state as $\hat{\bf{M}}(\pi/2,\pi)|1\rangle=\frac{1}{\sqrt{2}}\left(|1\rangle+|2\rangle\right)$ that accommodates a symmetric walk. We repeat the entire process, from creating a new BEC to applying as many step operators as necessary to implement each step of the walk. After each realization, the atomic distribution of the corresponding step is absorption imaged in a time-of-flight scheme, and finally all images are programmatically attached side-by-side to obtain the entire walk pattern. Our experimental observable is the momentum distribution, represented by the atomic population of momentum states $P(n)$ that we obtain from the walk images. Thus for an arbitrary state of the full system  $|{\psi}(j){\rangle}=\sum_{n}c_{n}(j)|n\rangle$, after the $j^{\rm{th}}$ step of the walk, we measure the momentum distribution as
$P(n,j)=P_{|1\rangle}(n,j)+P_{|2\rangle}(n,j)={|c_{n,1}(j)|}^2+{|c_{n,2}(j)|}^2$,
which contains the population distributions of both internal states. Experimentally, we capture both internal states in a single image by applying a short repump laser pulse to transfer the atoms at state $|1\rangle$ to state $|2\rangle$ before the imaging pulse.
In principle, in this type of experiment both the internal states could be measured independently, giving the two momentum distributions independently. This option may be interesting from a measurement theoretical perspective, to detect topological phases \cite{zhou2018floquet}, and in practice to give more options for a directional steering of the walk.

\section{RESULTS AND DISCUSSION}
\label{sec:results}

\begin{figure}
	\begin{center}
		\centering
		\includegraphics[scale=0.0478]{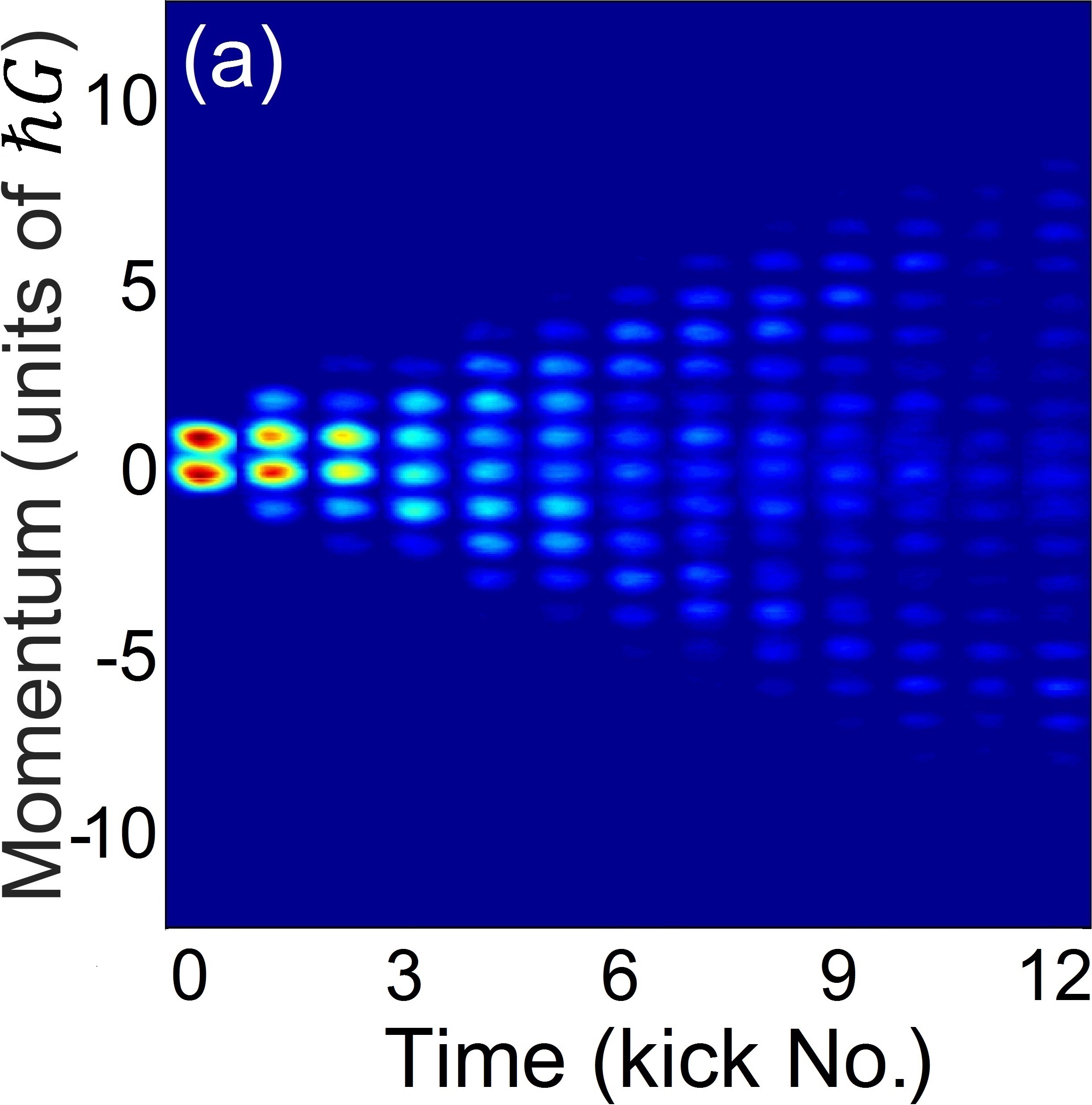}
		\includegraphics[scale= 0.0485]{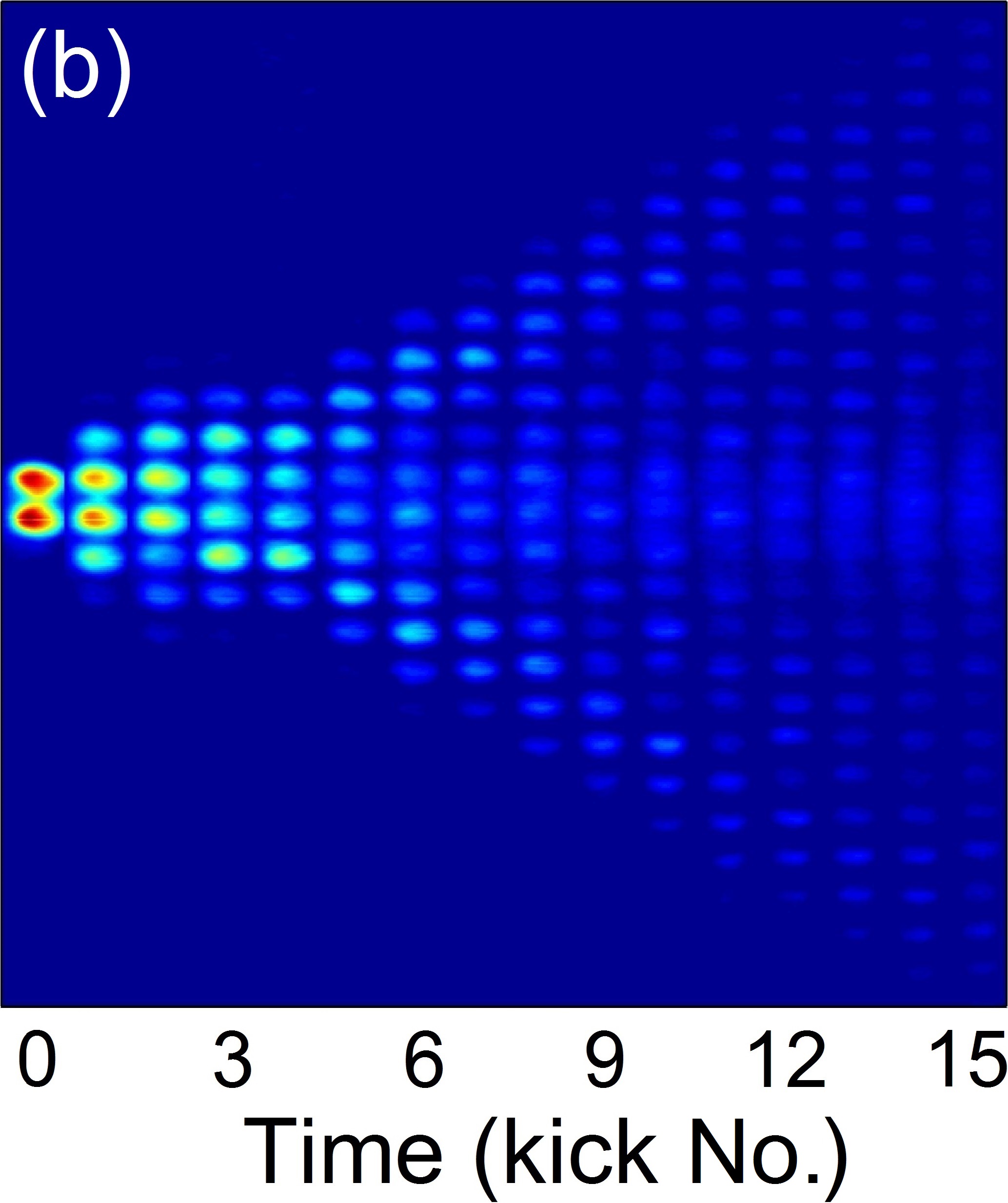}
		\includegraphics[scale=0.0480]{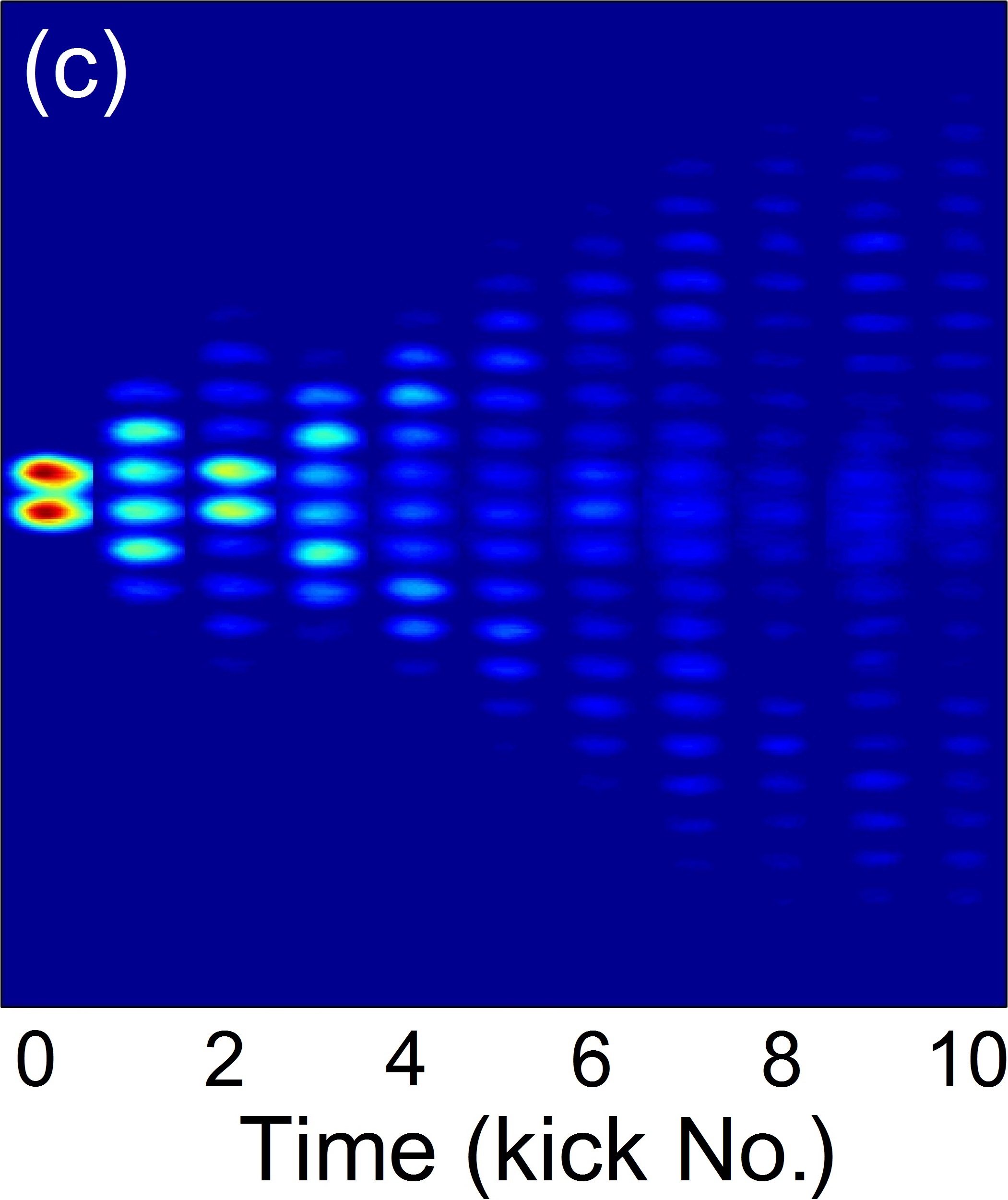}\\
		\vspace{0.3cm}
		\includegraphics[width=0.7\linewidth]{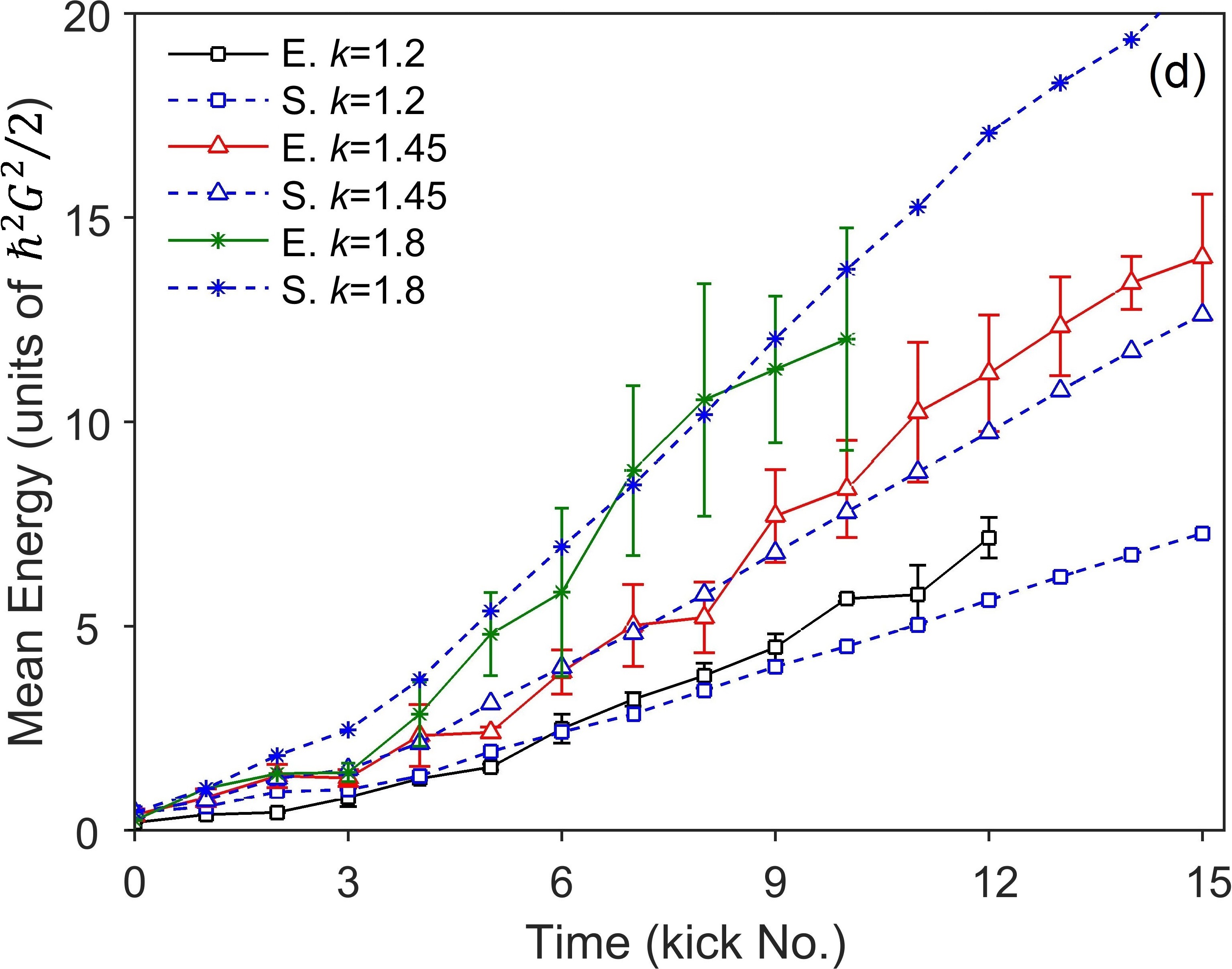}
\caption{Experimental momentum distributions of the standard QWs with ratchet strengths $|k|=1.2$ (a), 1.45 (b), and 1.8 (c). Each time (kick No.) represents one step of the walk, i.e one realization of the experiment. The case $|k|=1.45$ best matches the ideal QW by coupling neighboring momentum states. Panel (d) shows experimental (E.) and simulated (S.) mean energy growth of the walk vs. kicking strength. The walk with $|k|=1.8$ has only been performed up to 10 steps because its width is larger than the momentum detection window of the imaging system.} \label{fig2}
	\end{center}
\end{figure}

Figures~\ref{fig2}(a)-\ref{fig2}(c) show experimental results for the momentum distribution of the quantum walks realized with three different kick or ratchet strengths $|k|=1.2$, 1.45, and 1.8. These values are, in essence, the strengths of the AOKR, adjusted by varying the intensity of the optical standing wave. As can be seen, the characteristic standard deviation of the walk (spacing between the maxima) grows linearly with time (number of steps) for all three cases. However, as discussed in Sec. \ref{sec:QW}, the case $|k|=1.45$ best matches the ideal QW [with $q=1$ in Eq.~(\ref{eq2})] by coupling neighboring momentum states. Although the walk with $|k|=1.2$ is associated with less fluctuations in the momentum distribution and hence in the mean energy [see Fig.~\ref{fig2}(d)], the growth rate of these parameters are the lowest in this walk. On the other hand, the case $|k|=1.8$ shows a faster splitting of the walk peaks and hence a faster growth of momentum and energy. However, there are hints that in this case the walk has begun to suffer from decoherence and dephasing induced by the more intense light [note the saturation of the energy for this case in Fig.~\ref{fig2}(d)]. Thus, we employ the strength $|k|=1.45$ for investigating different aspects of our QWs in the subsequent experiments. We also draw attention to the fact that, regardless of the ratchet strength, all three walks have some diffusion between the diverging momentum peaks. However, since our shift operator works based on the quantum ratchet, we should be able to achieve walks with less diffusion through the manipulation of the initial state; by increasing the number of components contributing to the initial momentum state, we can achieve narrower spatial wave functions that result in ``cleaner'' ratchets \cite{ni2016initial,ni2017hamiltonian} and hence ``purer'' walk distributions.

\subsection{Effects of quasimomentum}
\label{sub1}

\begin{figure}
	\begin{center}
		\centering
		\includegraphics[scale=0.0485]{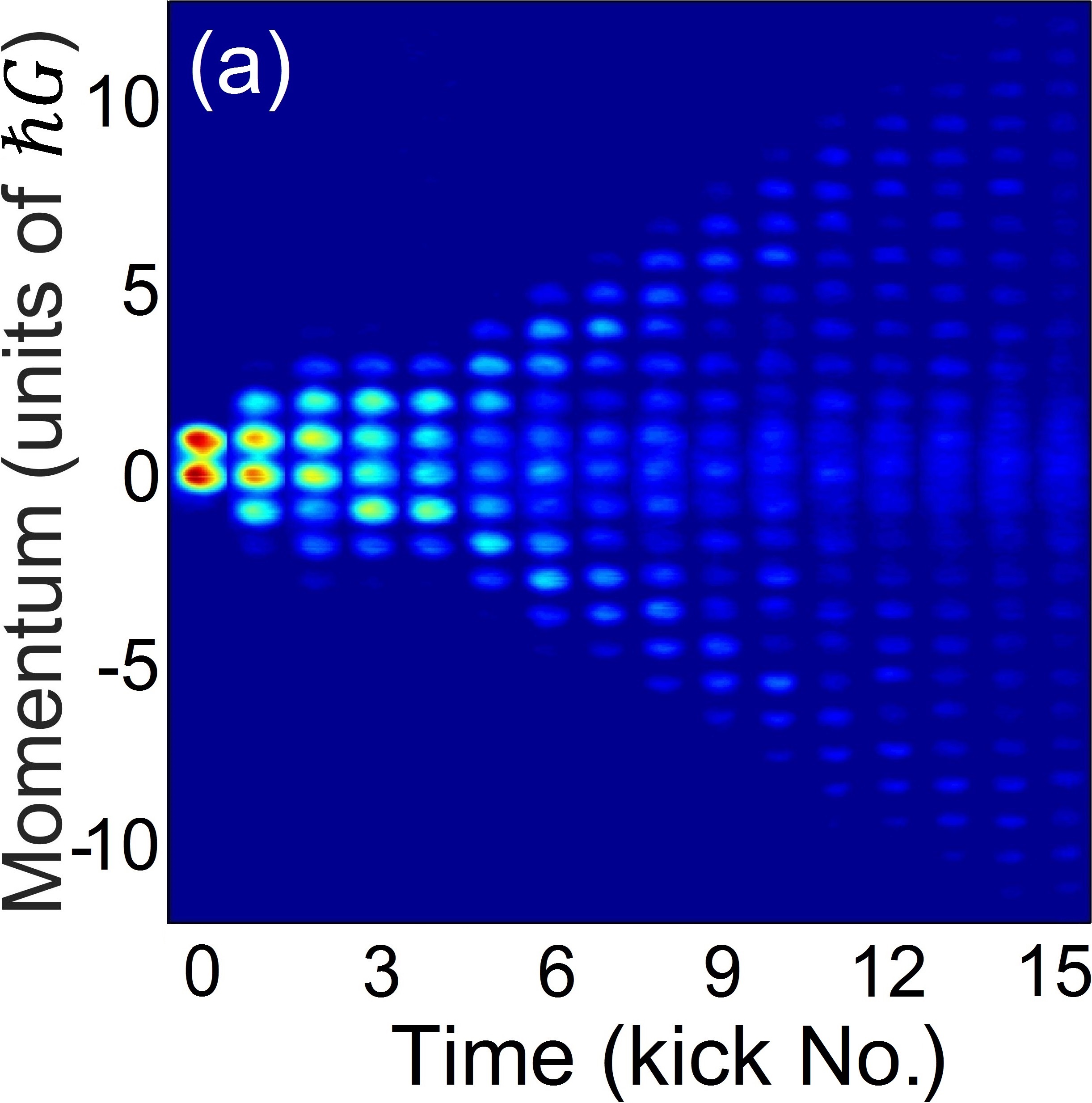}
		\includegraphics[scale= 0.0475]{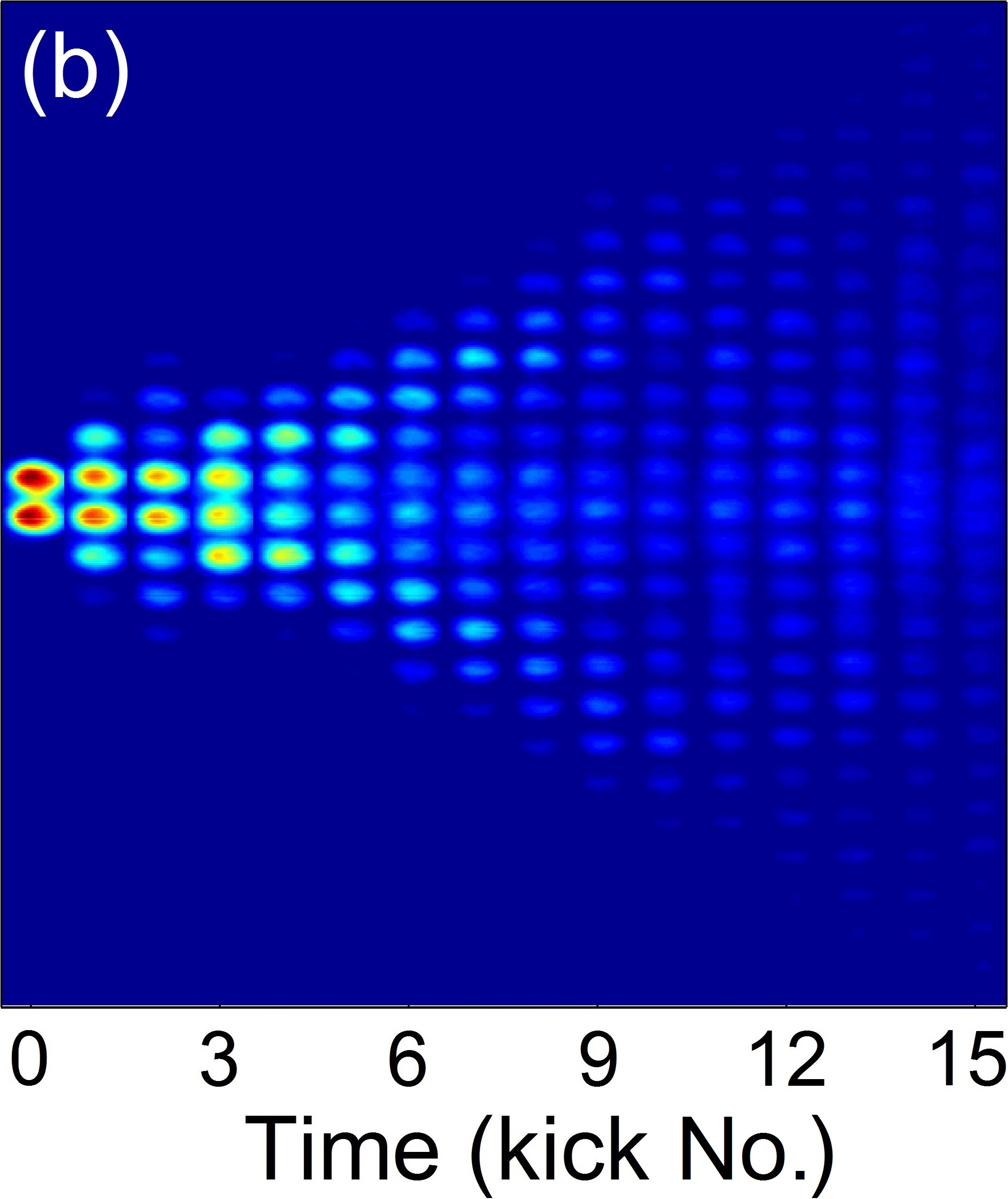}
		\includegraphics[scale=0.3095]{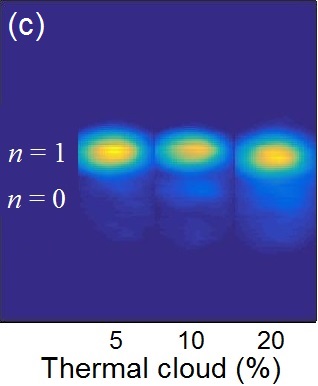}\\
		\vspace{0.3cm}
		\includegraphics[width=0.7\linewidth]{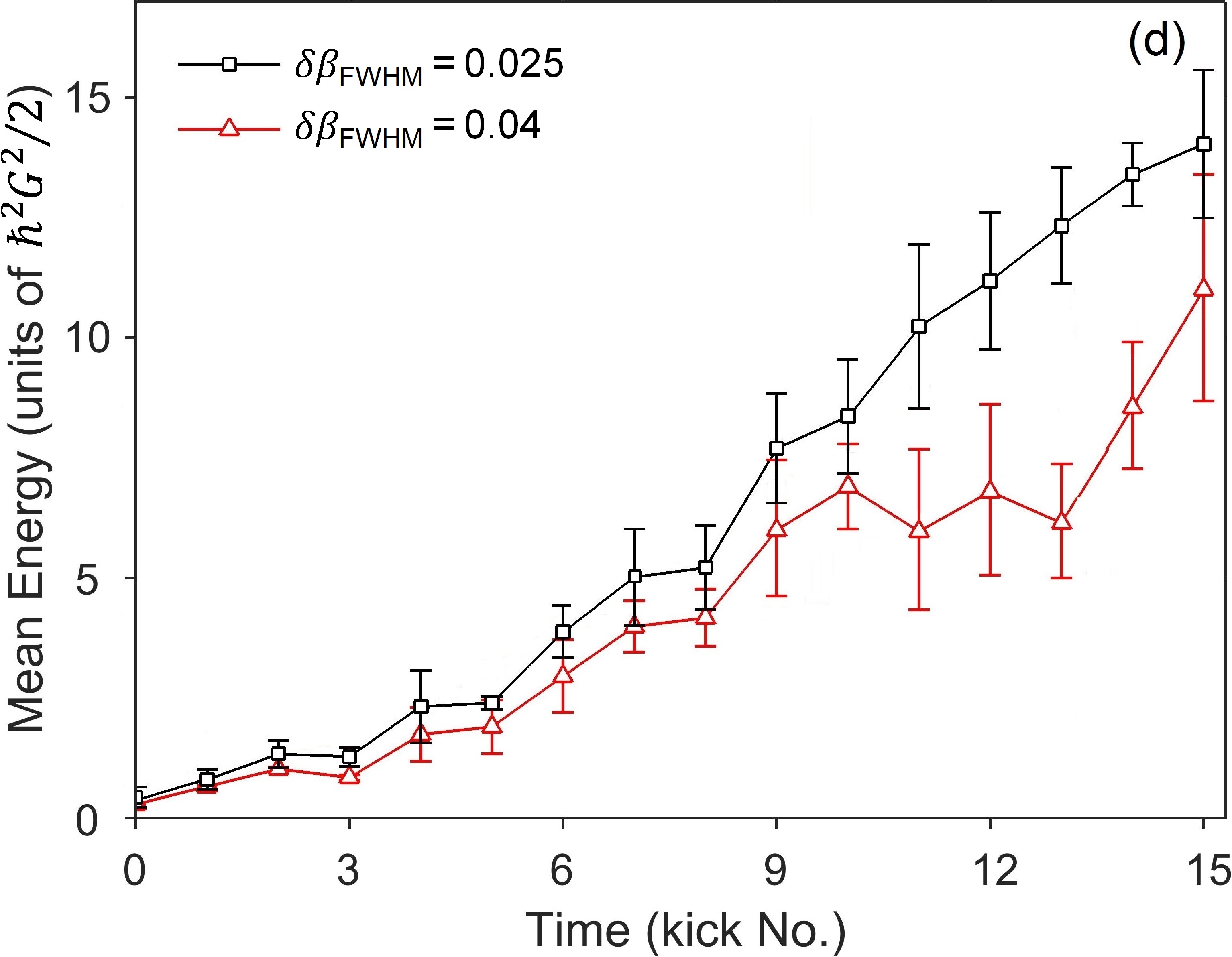}
\caption{Comparison of two standard QWs implemented using the BECs with different quasimomenta, $\delta\beta_{\mathrm{FWHM}}=0.025~\hbar G$ (a) and $0.04~\hbar G$ (b). The walk initiated with a larger quasimomentum undergoes a dephasing that prevents a significant fraction of the atomic population from being driven by the walk. Panel (c) demonstrates the effects of a strong Bragg pulse applied to an atomic population with different BEC/thermal cloud ratios. This pulse can almost entirely change the momentum state of the BEC from $|n=0\rangle$ to $|n=1\rangle$, whereas the thermal cloud, does not undergo this transition and remains mostly unaffected. Panel (d) shows the slower growth of the mean energy of the walk initiated with a BEC with larger quasimomentum.} \label{fig3}
	\end{center}
\end{figure}

In all of our QW images, there appears to be an ``amorphous'' population signal about the center of the momentum distribution. This is due to the near-resonant quasimomentum and a residual atomic thermal cloud that remains after the BEC is formed. In fact, a nonresonant quasimomentum can follow the resonant AOKR only for a few steps until the dephasing induced by the offset from the resonant values becomes strong enough to freeze its distribution into a broad Gaussian which is no longer affected by the ratchet process \cite{dadras2018discrete, Weiss2015, summy2016quantum}. By changing the evaporation sequence, we created atomic populations with varying amounts of thermal cloud. Figure~\ref{fig3} shows the adverse effects of increasing the quasimomentum on the momentum distribution [Figs.~\ref{fig3}(a) and \ref{fig3}(b)] and the corresponding mean energy of the walk [Fig.~\ref{fig3}(d)]. The thermal cloud includes a wide range of quasimomenta which are mostly off-resonant and thus do not respond to the optical pulses. We checked this point by applying a strong Bragg pulse on the BEC with various thermal cloud contributions. As shown in Fig.~\ref{fig3}(c), the Bragg pulse can entirely transform the BEC from the momentum state $|n=0\rangle$ to $|n=1\rangle$, but the thermal cloud does not contribute to the Bragg diffraction and remains unaffected.

Realizing the highest quality of a QW might encourage one to use a BEC with the minimum achievable quasimomentum and thermal cloud. However, this raises another difficulty. Experimentally, the walk is always associated with a gradual reduction of atomic population. Thus, using a BEC with these parameters minimized (which can be achieved only by significantly reducing the BEC population), results in a lack of visibility on the detection system and adversely affects the signal-to-noise ratio of the images. This trade-off limited us to realizing the optimum QWs with a $\delta\beta_{\mathrm{FWHM}}=0.025~\hbar G$ quasimomentum, corresponding to a $5--10\%$ thermal cloud of the atomic population.  

\subsection{Controlling the microwave phase}
\label{sub2}

\begin{figure}
	\begin{center}
		\centering
		\includegraphics[scale=0.097]{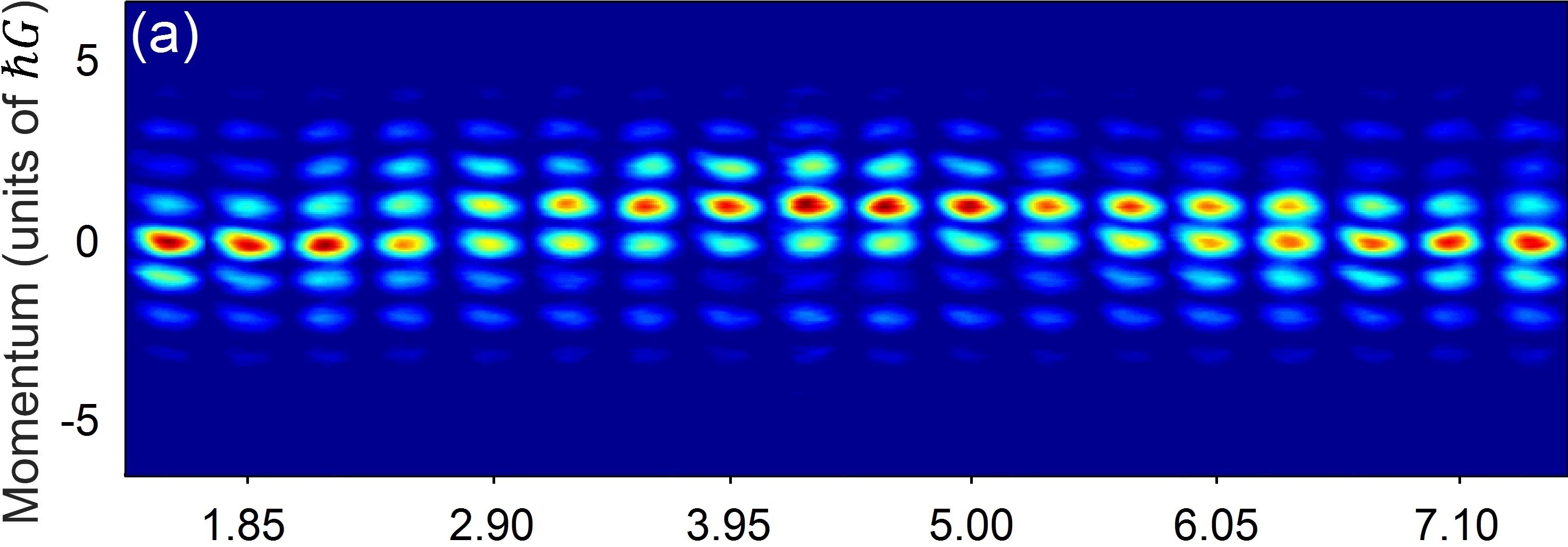}\\
		\vspace{0.1cm}
		\includegraphics[scale= 0.097]{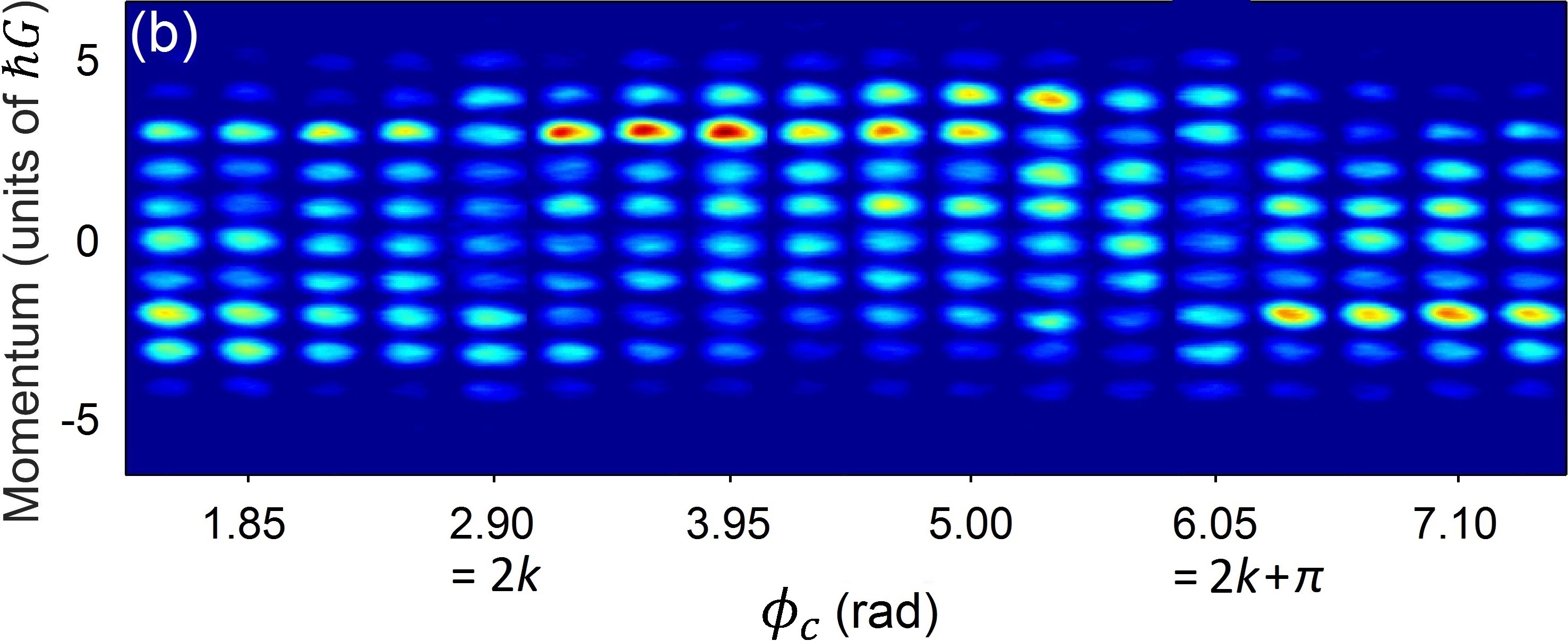}\\
		\vspace{0.3cm}
		\includegraphics[width=0.7\linewidth]{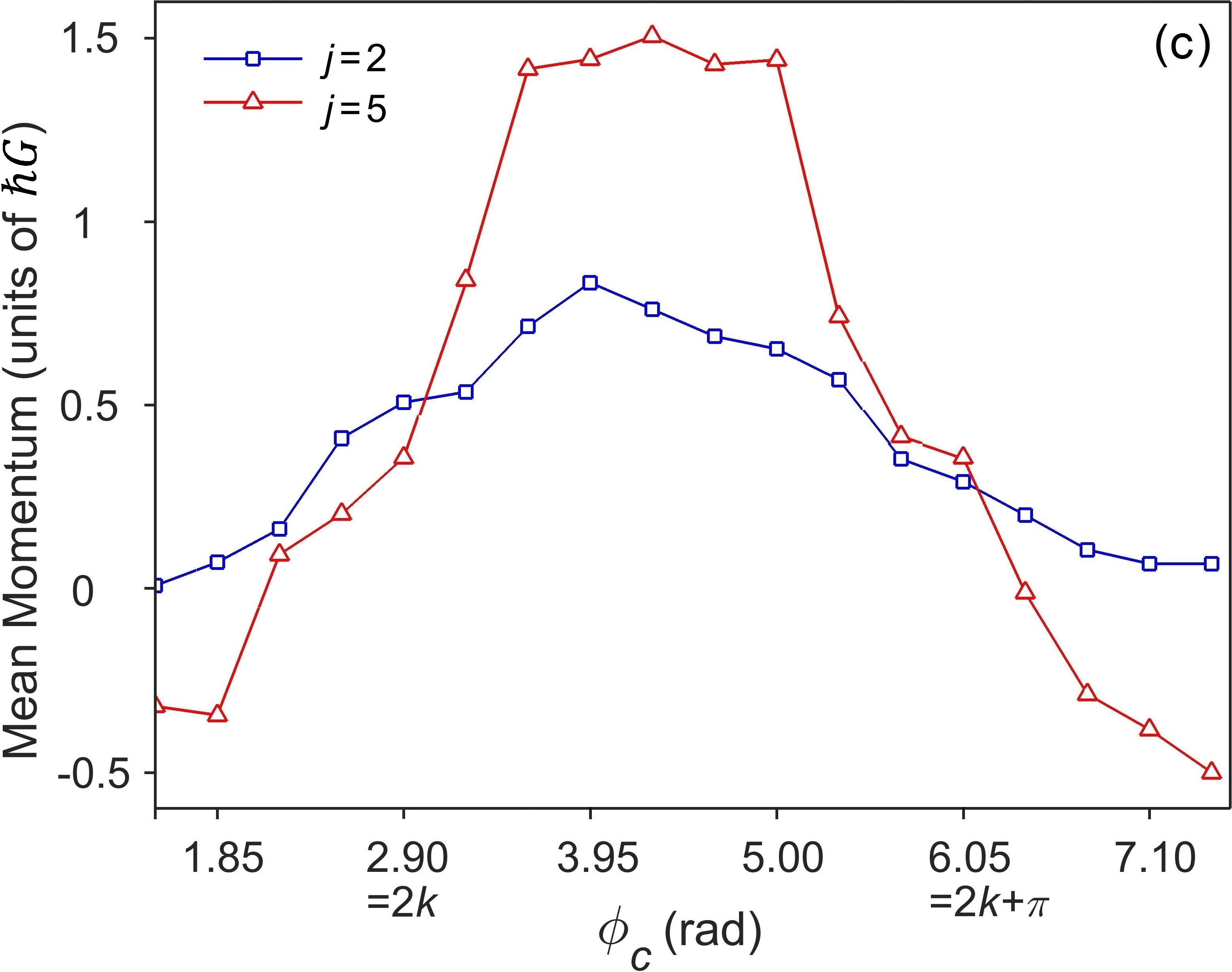}
		\caption{Scanning the phase $\phi_c$ of the MW pulses to verify the exact values required for suppressing the global phase effect due to the ratchet potentials. Panels (a) and (b) demonstrate the 2nd- and 5th-step momentum distributions vs $\phi_c$ and panel (c) shows the variation of their corresponding mean momenta. The oscillation of the weight of the walk by varying the phase is quite clear in these figures. As expected, the symmetry (mean momentum $\approx$0.5) is achieved only at $\phi_c=2k$ and $\phi_c=2k+\pi$~(rad).}	\label{fig4}
	\end{center}
\end{figure}

Another experimental concern that, if not addressed, can adversely affect our QWs is a ``global'' phase, introduced by the dc component of the kicking potential ${\hat{\mathcal{U}}}_k=e^{-ik[1+\cos(\hat{\theta})]}$. This is because the prefactor $e^{-ik}$, corresponding to this phase, can propagate through the quantum ratchet derivations \cite{dadras2018discrete} and consequently alter the ideal shift operator in Eq.~(\ref{eq3}) as
\begin{equation}
\hat{\bf{T}}=
\begin{bmatrix}
e^{-ik} &0\\
0 & e^{+ik}
\end{bmatrix}
\begin{bmatrix}
|n+1\rangle \langle n| &0\\
0 & |n-1\rangle \langle n|
\end{bmatrix}.
\label{eq5}
\end{equation}
This additional matrix imposes a phase difference of $\Delta \phi=2k$ between the internal states at each application of the ratchet. If this phase difference is not compensated for properly at each step, it can adversely affect the dynamics of the walk.
Experimentally, we address this by introducing compensation phases to each coin-toss MW pulse as ${\hat{\bf{M}}}(\pi/2,-\pi/2+\phi_c)$. Figure~\ref{fig4} shows the momentum distributions of a QW at the 2nd and 5th steps when this compensation phase is scanned over a $2\pi$ range. The periodic transfer of the weight of the walk from one side to the other by varying the $\phi_c$ is quite clear in these figures. As one would expect, the symmetry is achieved only when the phases $\phi_c=2k$ and $\phi_c=2k+\pi$~(rad) are applied on the MW pulses to properly cancel the global phase. Thus, effectively, we use the MW pulses ${\hat{\bf{M}}}(\pi/2,-\pi/2+2k)$ to maintain the quality of the walk at each step. 

\subsection{Initializing the internal state}
\label{sub3}

\begin{figure}
	\begin{center}
		\centering
		\includegraphics[scale=0.167]{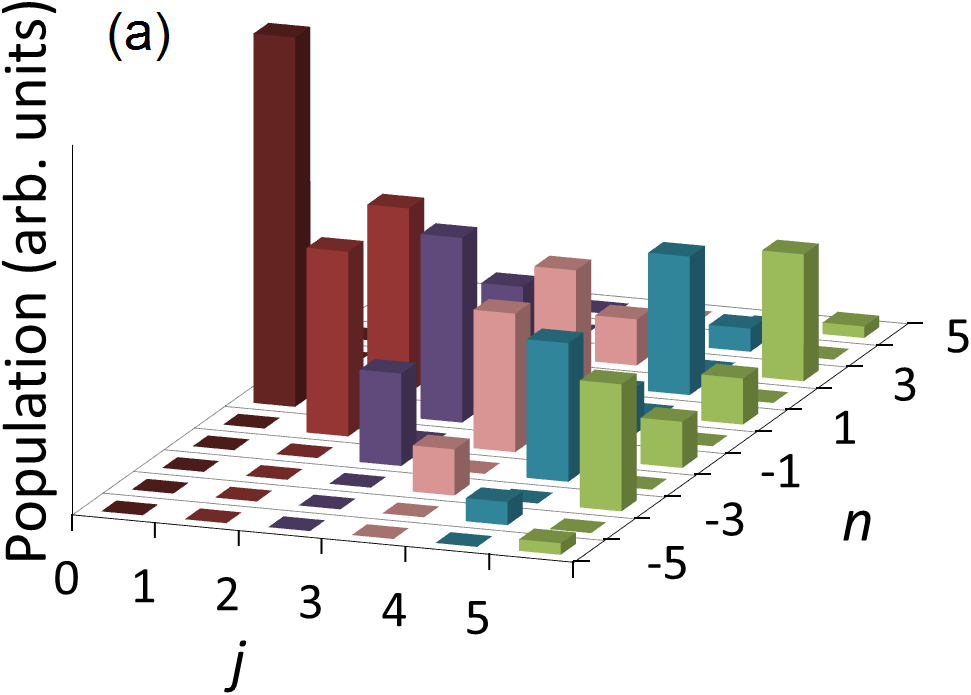}
		\hspace{0.01cm}
		\includegraphics[scale= 0.167]{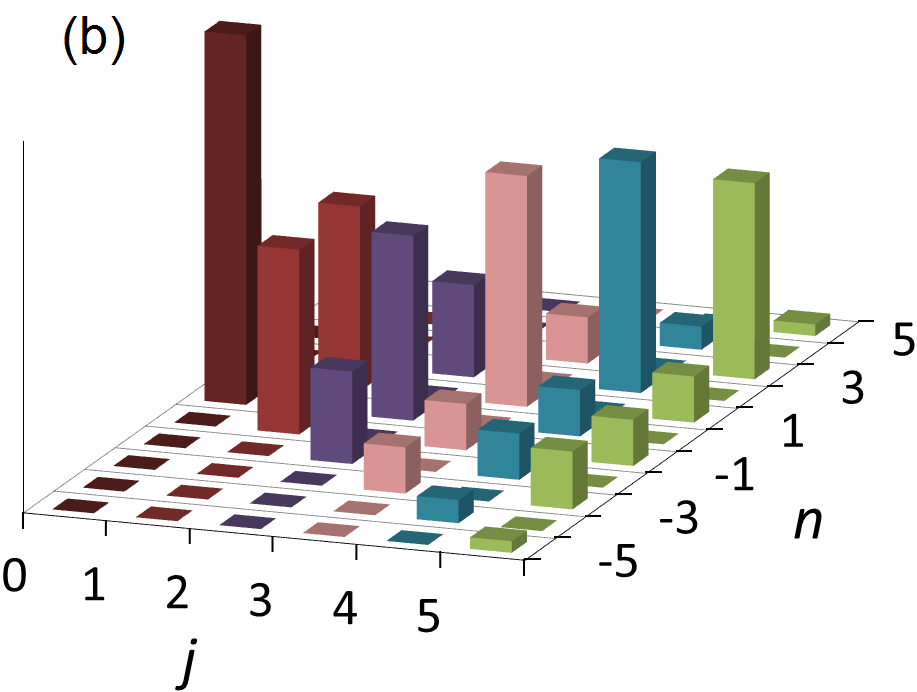}\\
		\caption{The population distributions of the first five steps of a QW when a gate operator different from (a) or similar to (b) the coin operators is used to initialize the walk; see main text for details. The walk starts to be asymmetric at the third step when the latter case is implemented.}	\label{fig5}
	\end{center}
\end{figure}

\begin{figure}
	\begin{center}
		\centering
		\includegraphics[scale=0.06]{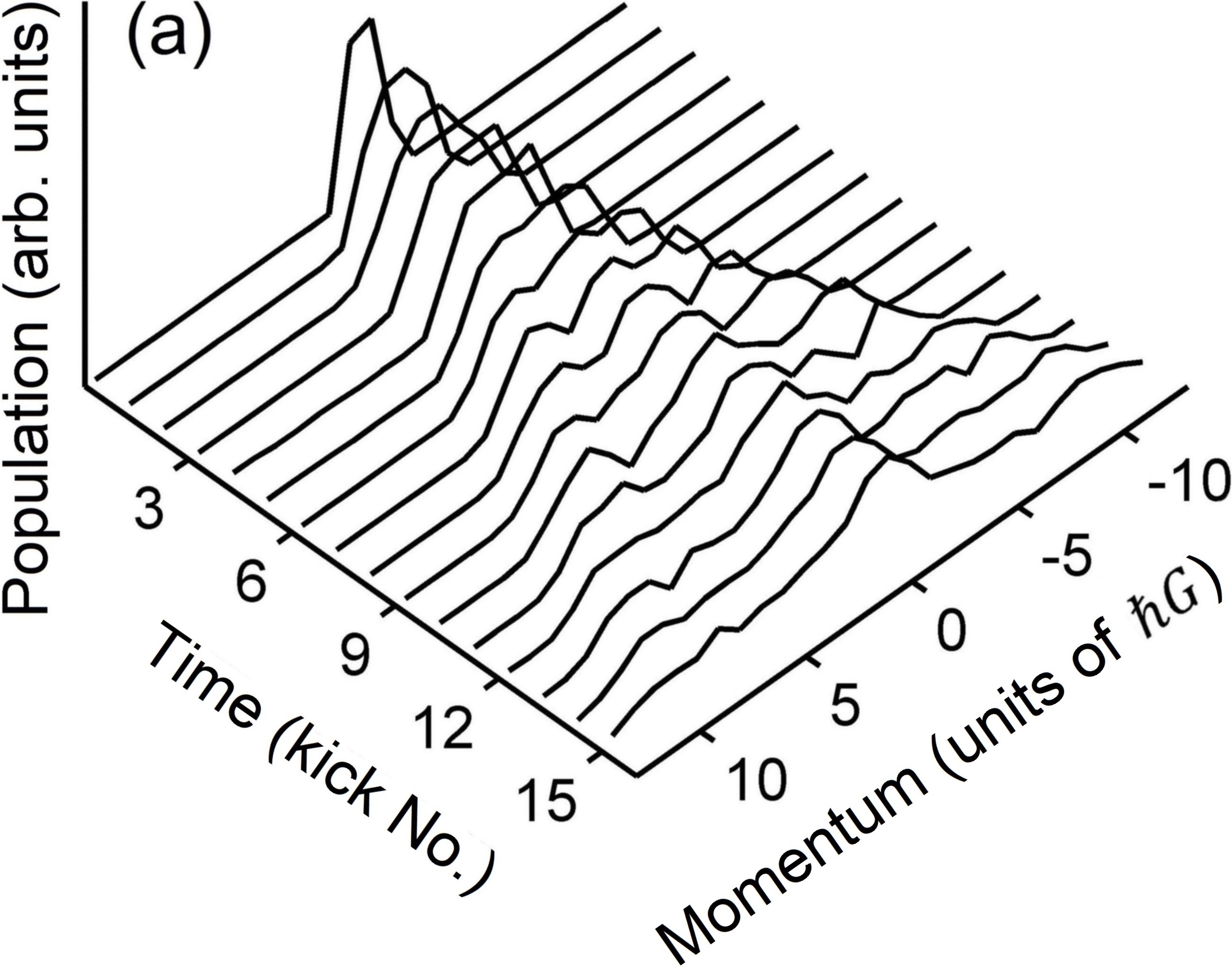}
		\hspace{0.1cm}
		\includegraphics[scale= 0.06]{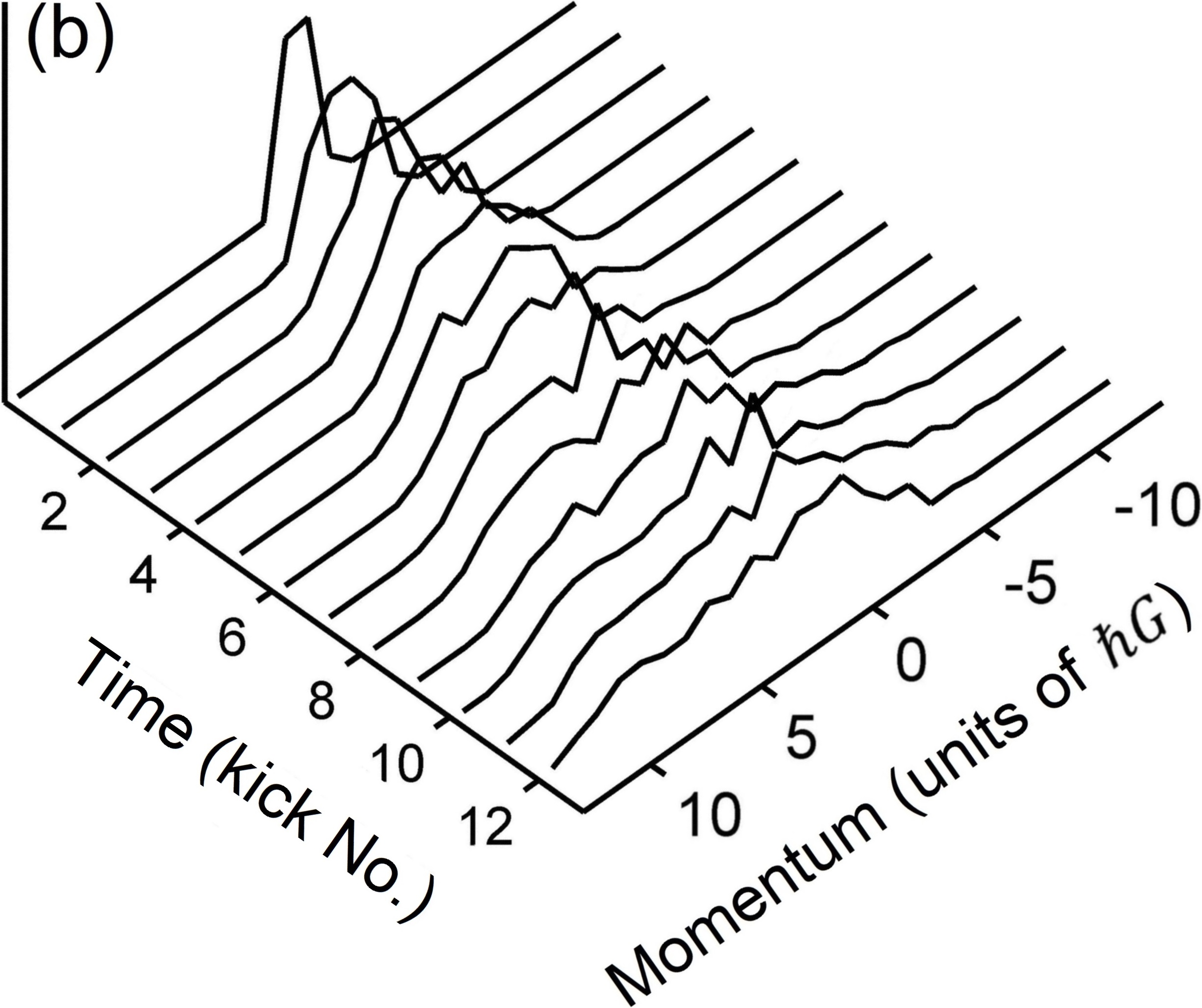}\\
		\vspace{0.3cm}
		\includegraphics[width=0.7\linewidth]{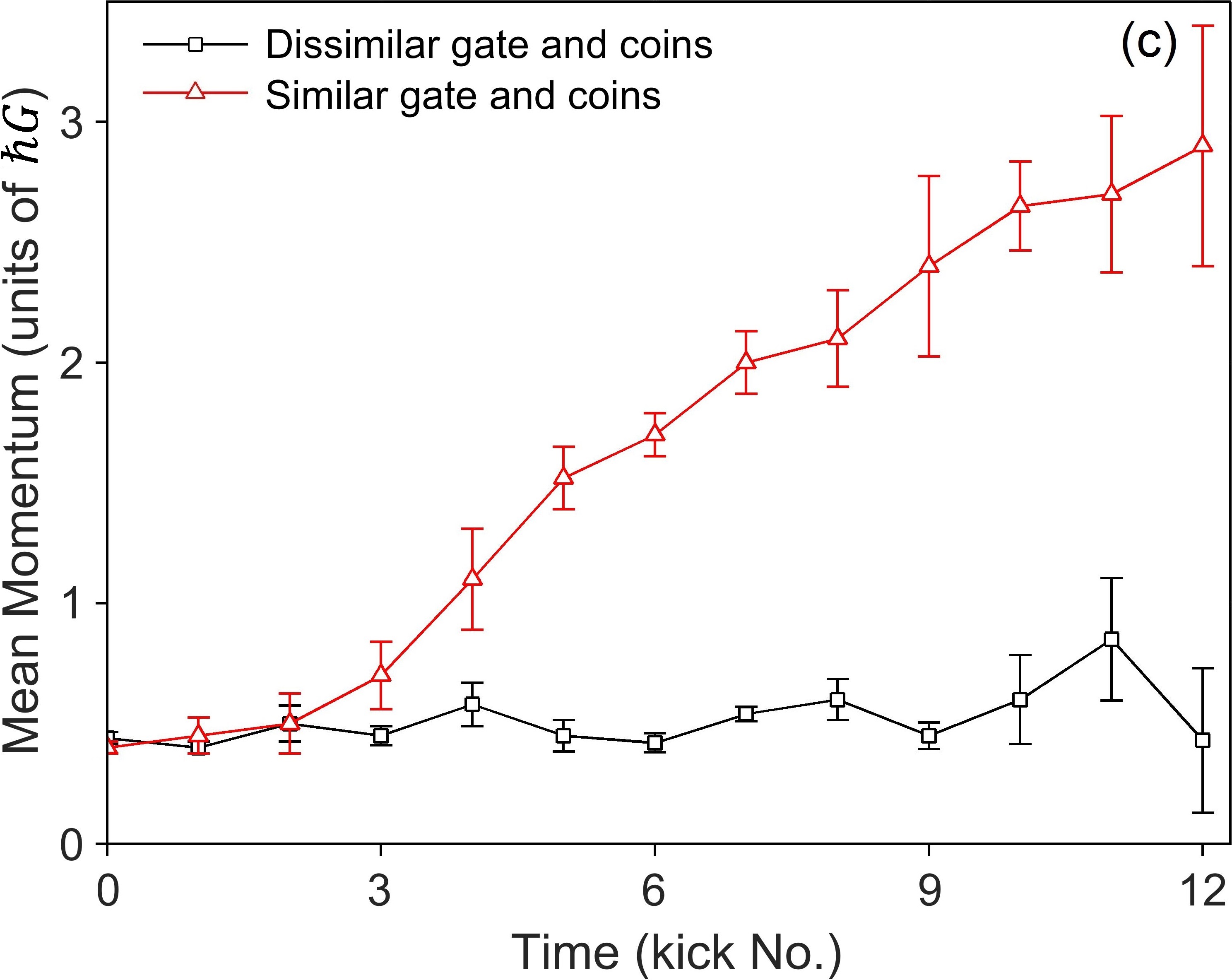}
		\caption{Experimental QWs implemented with dissimilar (a) and similar (b) gate and coin operators. The walk becomes asymmetric in the latter, leading to an increasing mean momentum as shown in panel (c).}\label{fig6}
	\end{center}
\end{figure}

In addition, the phase arrangement of the MW pulses used for initializing the walk (gate) and mixing the internal states at subsequent steps (coins) can affect the symmetry of the walk. This is due to the fact that the direction of the shifts in the external state is entangled to the walker's internal state, determined by a MW pulse at each step. A standard (symmetric) walk can only be realized with the choice of gate and coin operators with phase differences of odd multiples of $\pi/2$, e.g., $\hat{\bf{M}}(\pi/2,\pi)$ and $\hat{\bf{M}}(\pi/2,-\pi/2)$ for the Hadamard gate and coin operators. One can verify this by successively applying these operators on an initial state $|\Psi_0\rangle=|1\rangle\otimes|n=0\rangle$ for a few steps of the sequence mentioned in Sec. \ref{sec:exp.} (for simplicity we assume that the ratchet can still be achieved with one momentum state $|n=0\rangle$). Figure~\ref{fig5}(a) shows the symmetry in the atomic population of the momentum states $P(n)$, obtained for the first five steps of this walk. However, the symmetry starts to break quickly in the case of applying other configurations of MW pulses, e.g., using similar $\hat{\bf{M}}(\pi/2,-\pi/2)$ pulses for both gate and coin operations [see Fig.~\ref{fig5}(b)]. Figure~\ref{fig6} compares the experimental QW patterns obtained for these two different arrangements of the gate and coin pulses. 

\subsection{Quantum-to-classical transition}
\label{sub4}

\begin{figure}
	\begin{center}
		\centering
		\includegraphics[scale=0.0484]{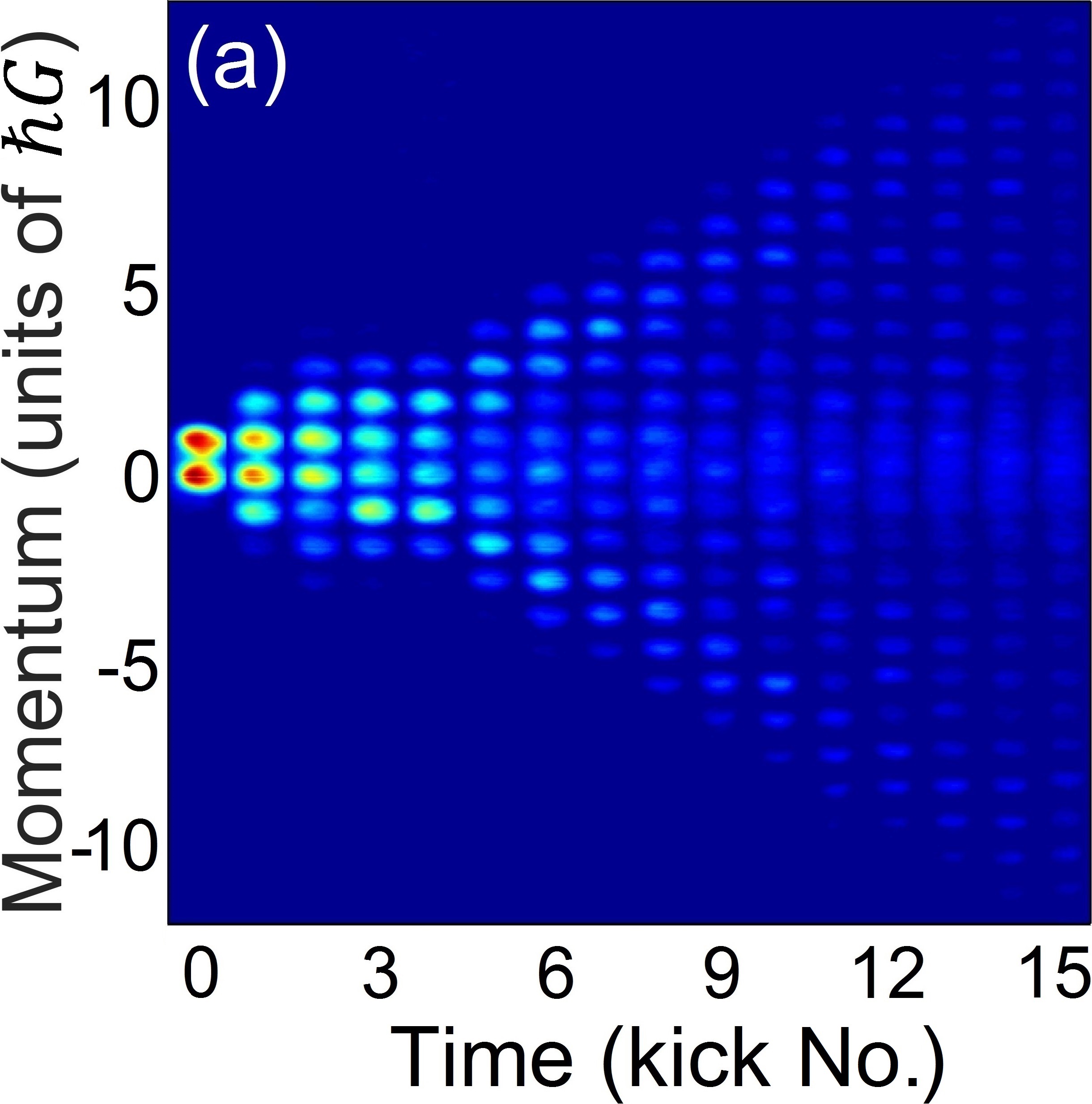}
		\includegraphics[scale= 0.0477]{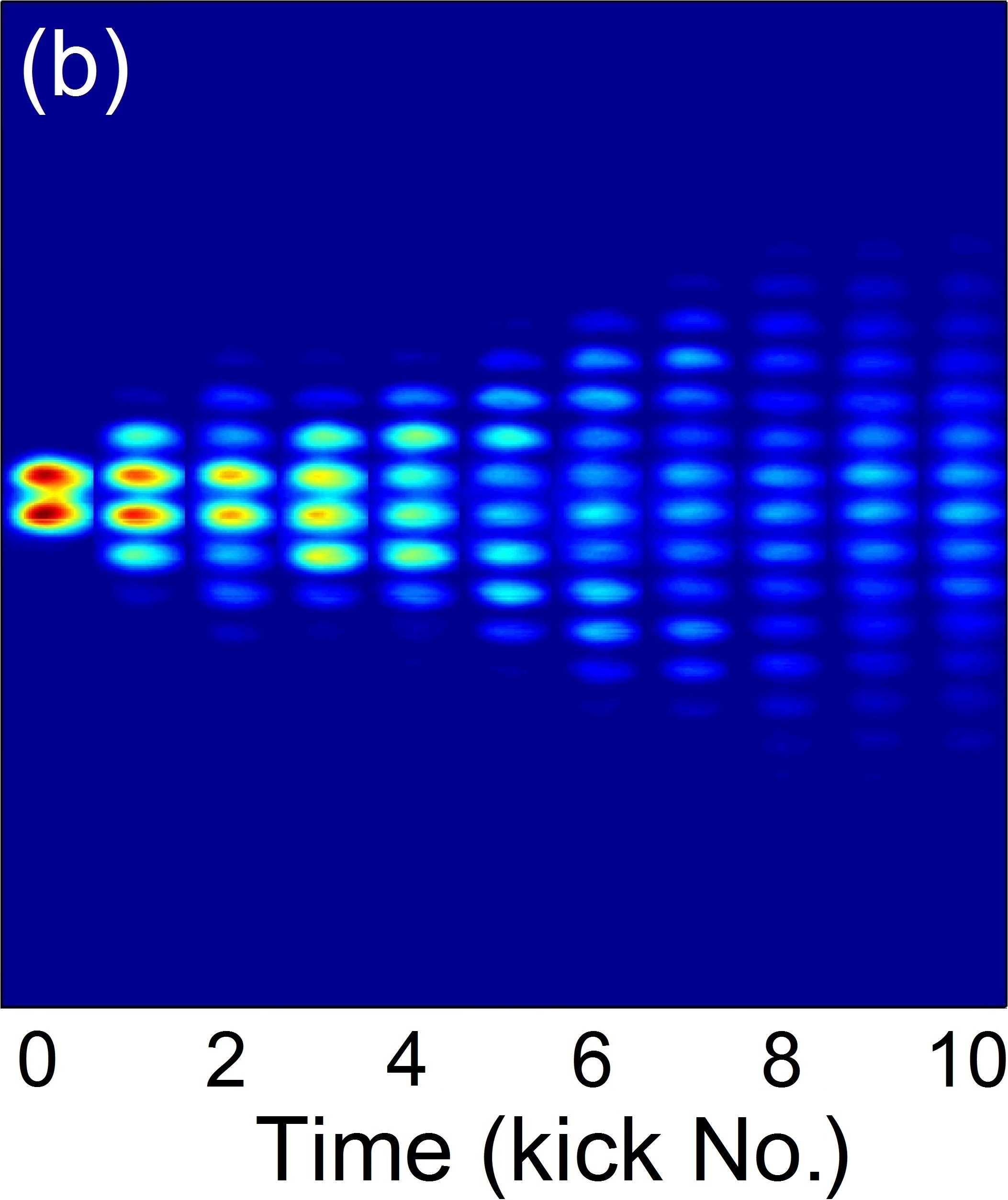}
		\includegraphics[scale=0.0477]{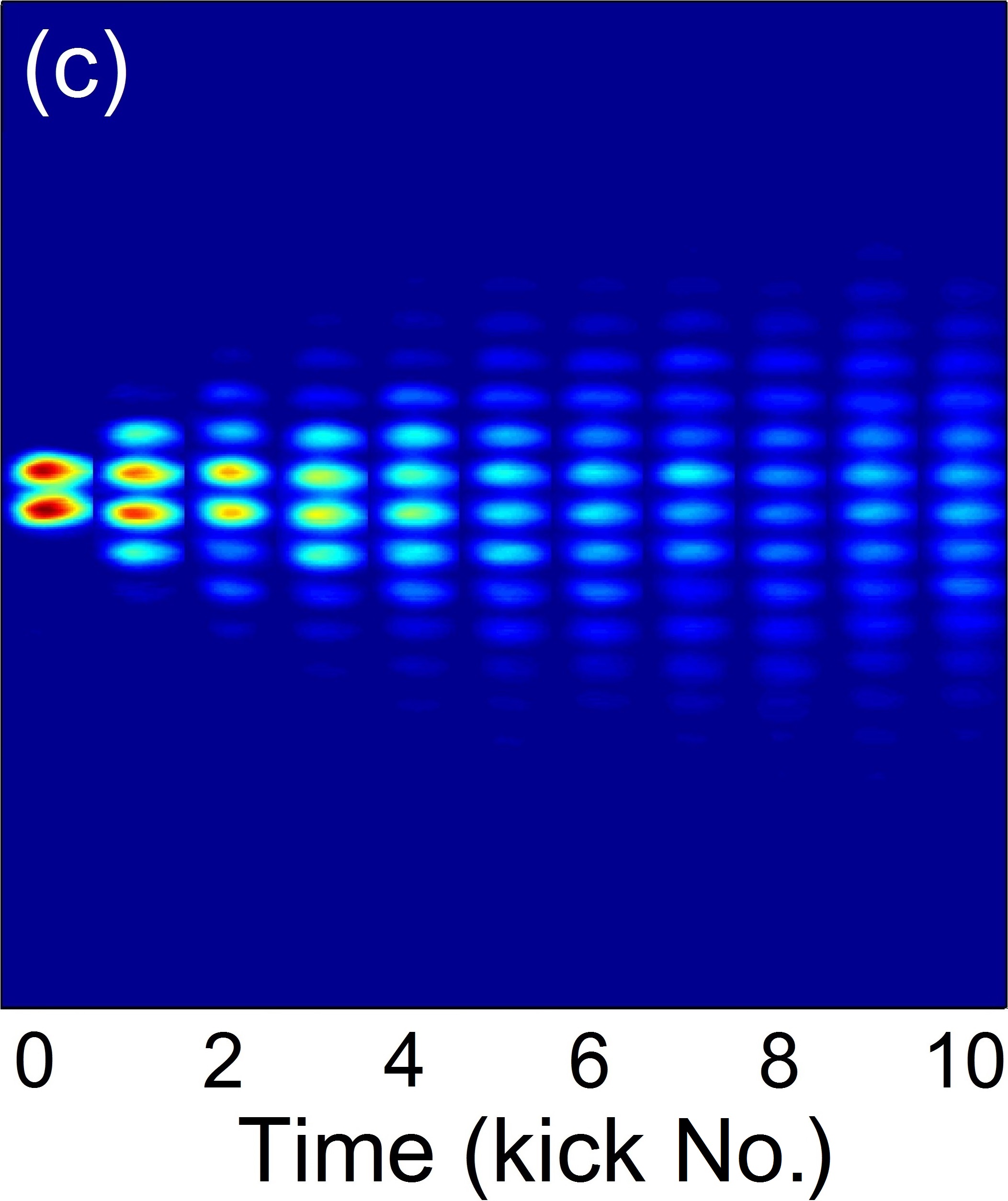}\\
		\vspace{0.3cm}
		\includegraphics[width=0.7\linewidth]{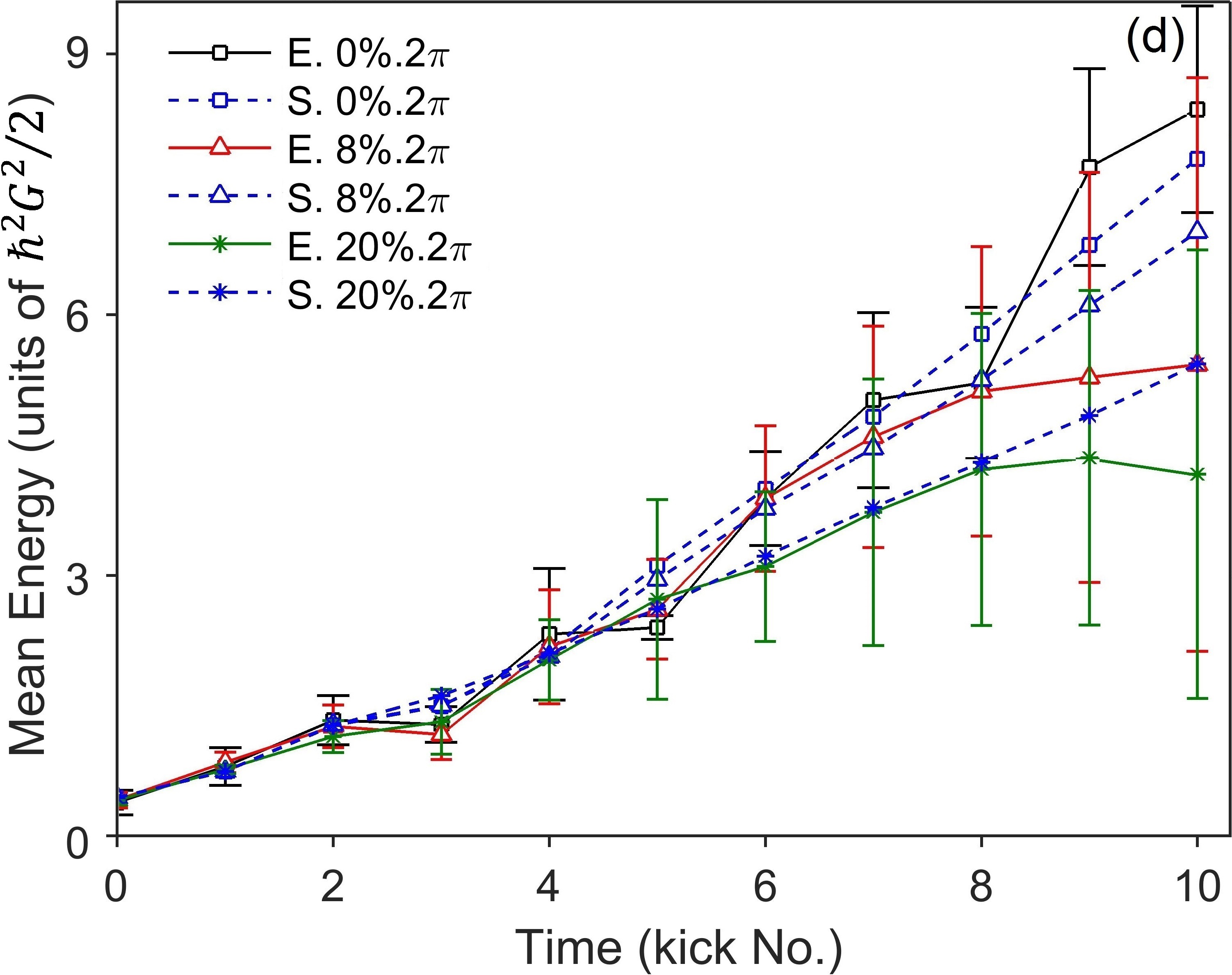}
		\caption{Transition of our QW to classical as a result of a noise enhancement. This noise takes the form of a controllable randomization of the phase of each MW pulse. The standard QW (a) was carried out with keeping the coin phase fixed at $\chi=\pi/2$. Signatures of a classical walk start to appear at $8\%$ phase randomness (b), and the walk becomes almost completely classical when the phase is randomized by $20\%$ (c). The corresponding mean energy variations (d) reveal a slowdown in the propagation of the walk as a result of this quantum-to-classical transition.}\label{fig7}
	\end{center}
\end{figure}

The transition to a classical walk is one of the interesting features that we are able to demonstrate in our experiments thanks to the relatively large range of our walk steps. We investigate this feature by randomizing the mixing between the two internal states during each coin toss by adding a uniformly-distributed random phase to the MW pulses. Figures~\ref{fig7}(a)-\ref{fig7}(c) demonstrate the ``snapshots'' of the quantum-to-classical transition of our walk, realized at different amounts of coin-phase randomness; Fig.~\ref{fig7}(a) represents the previously introduced standard QW [Fig.~\ref{fig1}(b)] with the characteristic standard deviation of the momentum distribution ${\propto}j$. This was conducted by keeping the coin phases fixed at $\chi=-\pi/2$. The QW ballistic peaks become less prominent after a few steps by adding as little as $8\%$ randomness, Fig.~\ref{fig7}(b). The disappearance of these peaks, along with the emergence of a Gaussian distribution with the characteristic standard deviation ${\propto}\sqrt{j}$, is a manifestation of a classical walk. The walk becomes dominantly classical (Gaussian with no QW peaks) at $20\%$ phase randomness [Fig.~\ref{fig7}(c)] and fully classical by gradually increasing the noise within a full $2\pi$ (not shown here). Figure~\ref{fig7}(d) shows the behavior of the mean energy of these walks. As can be seen, with larger amounts of phase randomness the mean energy of the system grows at a slower rate and displays characteristics of a classical walk. 

\subsection{Biased walks}
\label{sub5}

\begin{figure}
	\begin{center}
		\centering
		\includegraphics[scale=0.15]{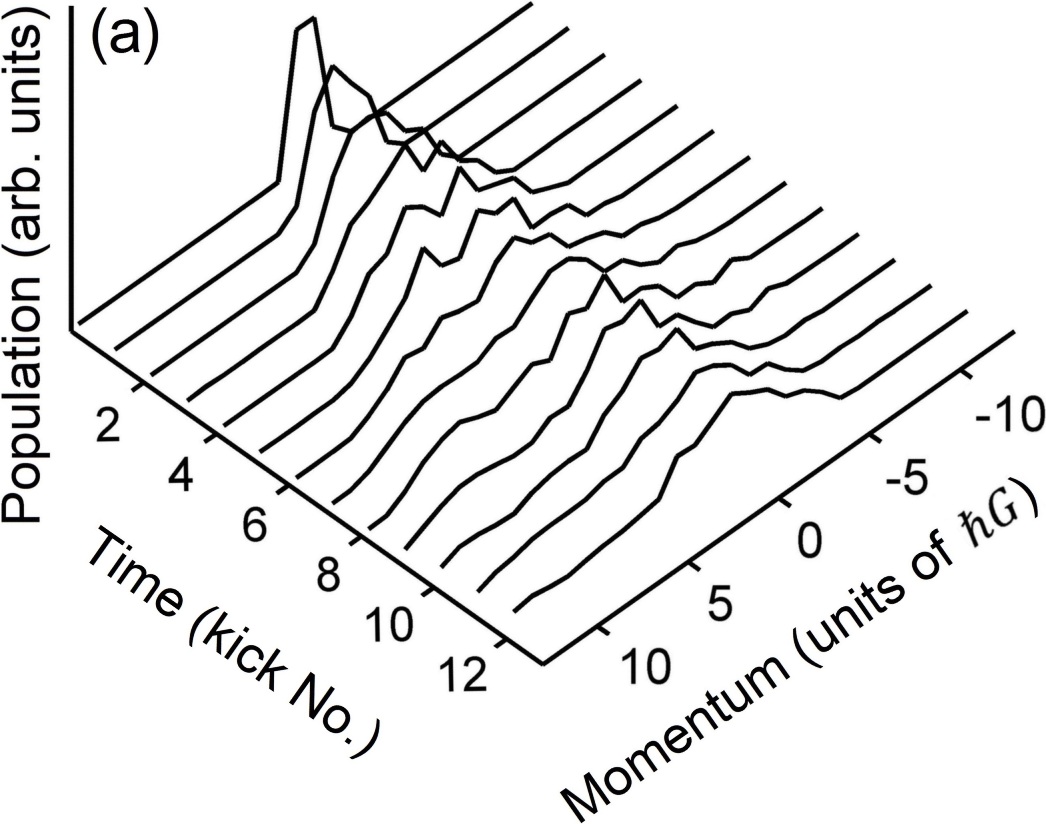}
		\hspace{0.1cm}
		\includegraphics[scale= 0.15]{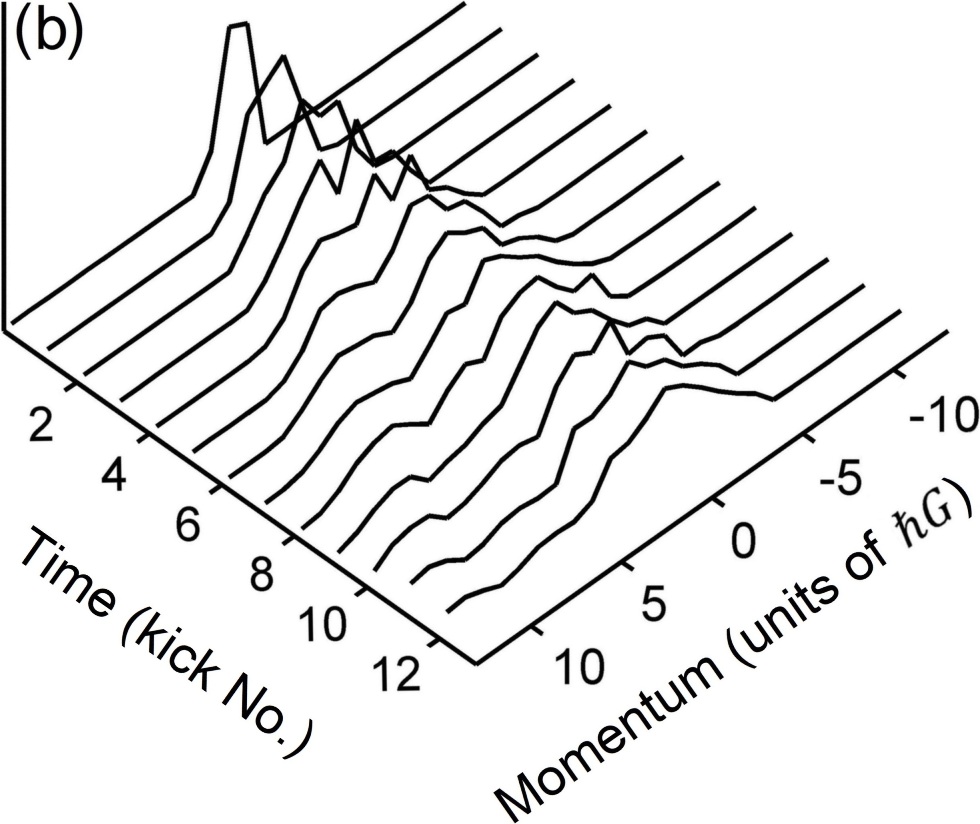}\\
		\vspace{0.3cm}
		\includegraphics[width=0.7\linewidth]{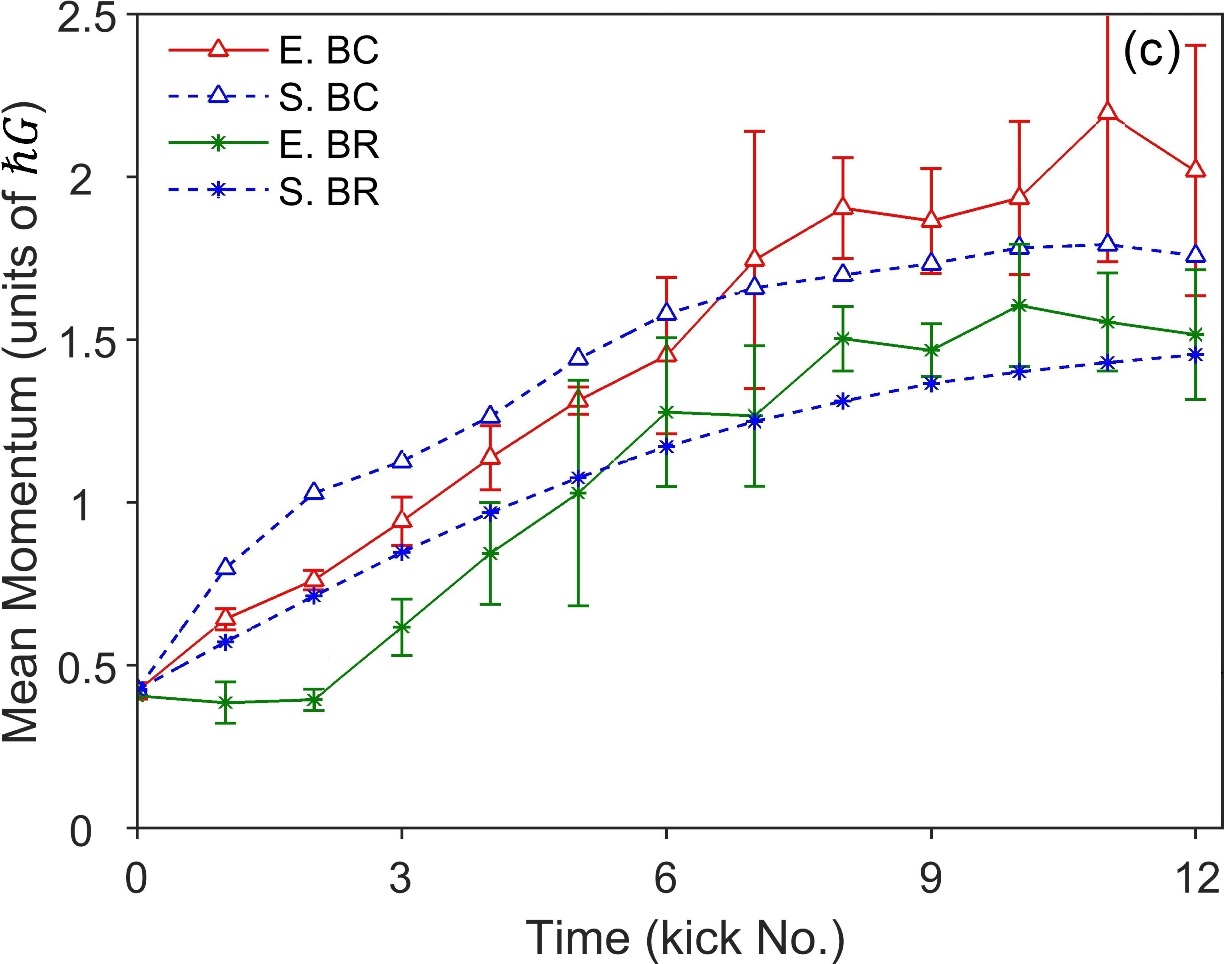}
		\caption{Quantum walks steered using the BCs with $\rho=0.7$ and $|k|=1.45$ (a), and BRs with $k_{1}=-1.7$, $k_{2}=+1.0$ (b). The former produces an unequal superposition of internal states at each coin toss, and the latter applies unequal shifts on the momentum of each internal state. Panel (c) shows the experimental (E.) and simulated (S.) variations of the mean momenta of these walks. Compare these steered walks and their mean momenta to those of a symmetric walk [Figures~\ref{fig6}(a) and \ref{fig6}(c) for dissimilar gate and coins] implemented with $\rho=0.5$ and $k_{1}=-1.45$, $k_{2}=+1.45$.} \label{fig8}
	\end{center}
\end{figure}

We are also able to steer our QWs by controllably biasing either coin or shift operators, i.e., by manipulating either internal or external degrees of freedom.
In the former, we realize the biased coins (BCs) by altering the power of the MW pulses from the ${\pi}/2$ scheme to obtain unequal superpositions of internal states. This effectively breaks the balance of our standard gate and coin operators by introducing a bias factor $\rho$ such that \cite{dadras2018discrete}
\begin{align}
\hat{\bf{M}}_\rho(\pi)=
\begin{pmatrix}
\sqrt{\rho} & -\sqrt{1-\rho}\\
\sqrt{1-\rho} & \sqrt{\rho}
\end{pmatrix}, \hspace{0.5cm}\nonumber\\
\hspace{-1.4cm}\mathrm{and,}\hspace{0.7cm}\hat{\bf{M}}_\rho(-\pi/2)=
\begin{pmatrix}
\sqrt{\rho} & i\cdot\sqrt{1-\rho}\\
i\cdot\sqrt{1-\rho} & \sqrt{\rho}
\end{pmatrix}.
\label{eq:bgate}
\end{align}
Applying these biased operators at each step constantly assigns a higher population to one internal state than the other. Figure~\ref{fig8}(a) demonstrates the experimental results of a BC walk implemented with $\rho=0.7$ [compare to Fig.~\ref{fig6}(a) for an unbiased walk realized with $\rho=0.5$ and the same ratchet strength $|k|=1.45$]. In the second form of the steered walks, a biased ratchet (BR) is applied to realize non-symmetric walk steps. We achieve the BR by detuning the kicking laser so that the laser frequency is no longer halfway between the ground-state hyperfine levels. This shift in the laser frequency results in unequal ratchet potentials addressing each state [see definition of $k$ around Eq.~(\ref{eq4})] and generalizes the original shift operator $\hat{\bf{T}}=e^{-ik\cos(\hat{\theta}).\sigma_z}$ to \cite{summy2016quantum}
\begin{equation}
\hat{\bf{T}}=\exp\left[-i\cos(\hat{\theta})
\begin{pmatrix}
k_1 &0\\
0 & k_2
\end{pmatrix}\right],
\label{eq:bratchet} 
\end{equation}
where the bias is controlled by $k_2/k_1$. Figure~\ref{fig8}(b) shows a BR steered walk realized with the kicking strengths $k_{1}=-1.7$, $k_{2}=+1.0$. The corresponding mean momenta of these BC and BR steered walks are also demonstrated in Fig.~\ref{fig8}(c), displaying their apparent growth compared to the unvarying momentum of the symmetric walk shown in Fig.~\ref{fig6}(c). Intuitively, by adjusting the coin and ratchet bias factors, one can engineer the direction and speed of the walk as desired or compensate for any probable biases in the walk.

\subsection{Time reversal}
\label{sub6}

\begin{figure}
	\begin{center}
		\centering
		\includegraphics[scale=0.145]{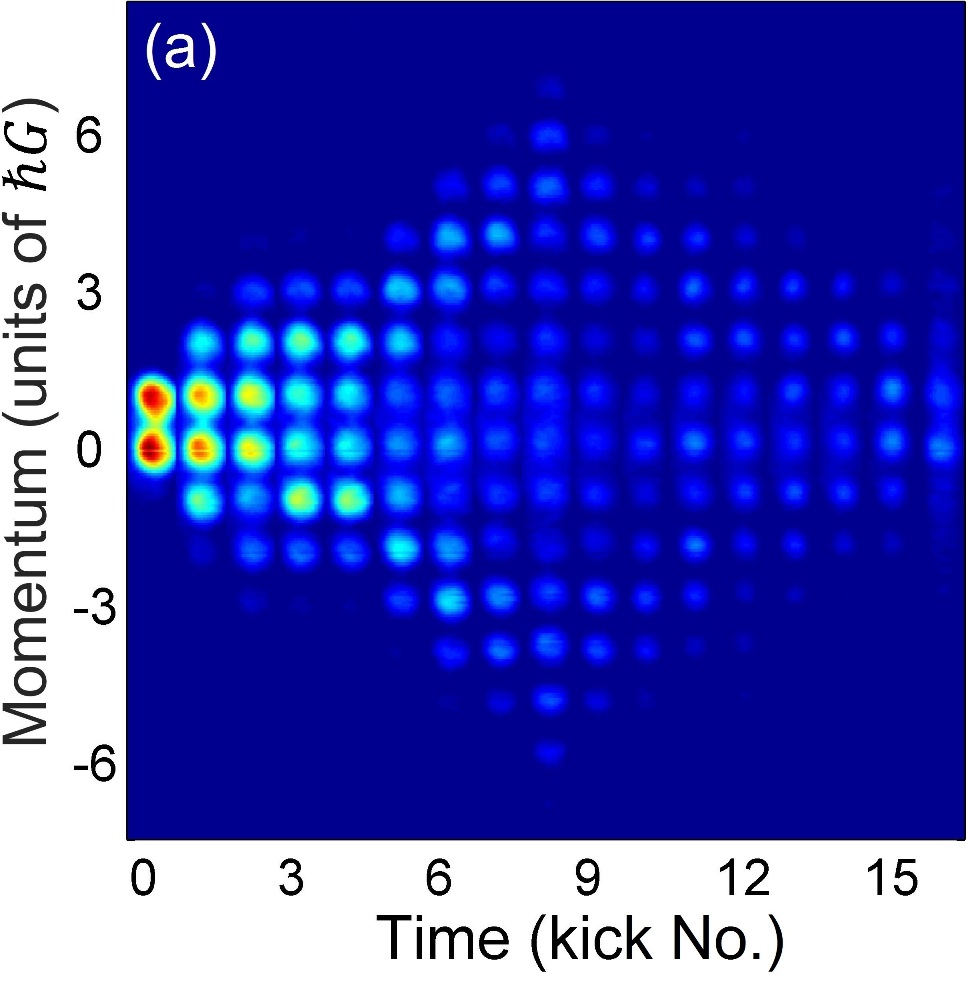}
		\hspace{0.1cm}
		\includegraphics[scale= 0.17]{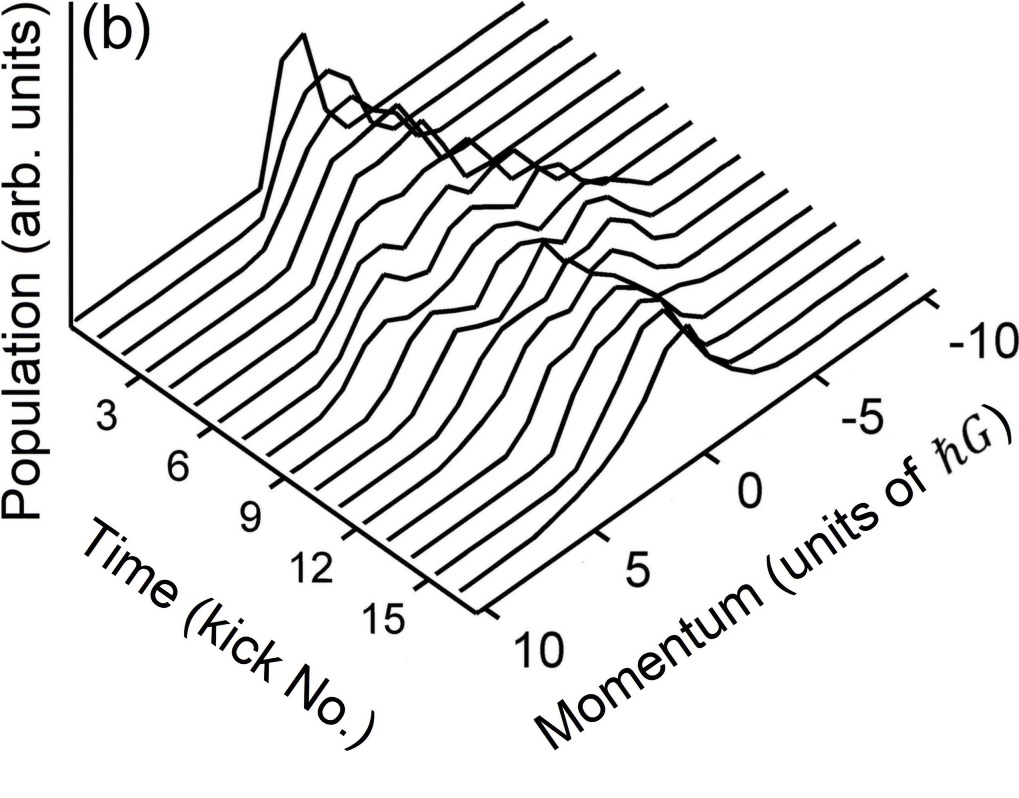}\\
		\vspace{0.3cm}
		\includegraphics[width=0.7\linewidth]{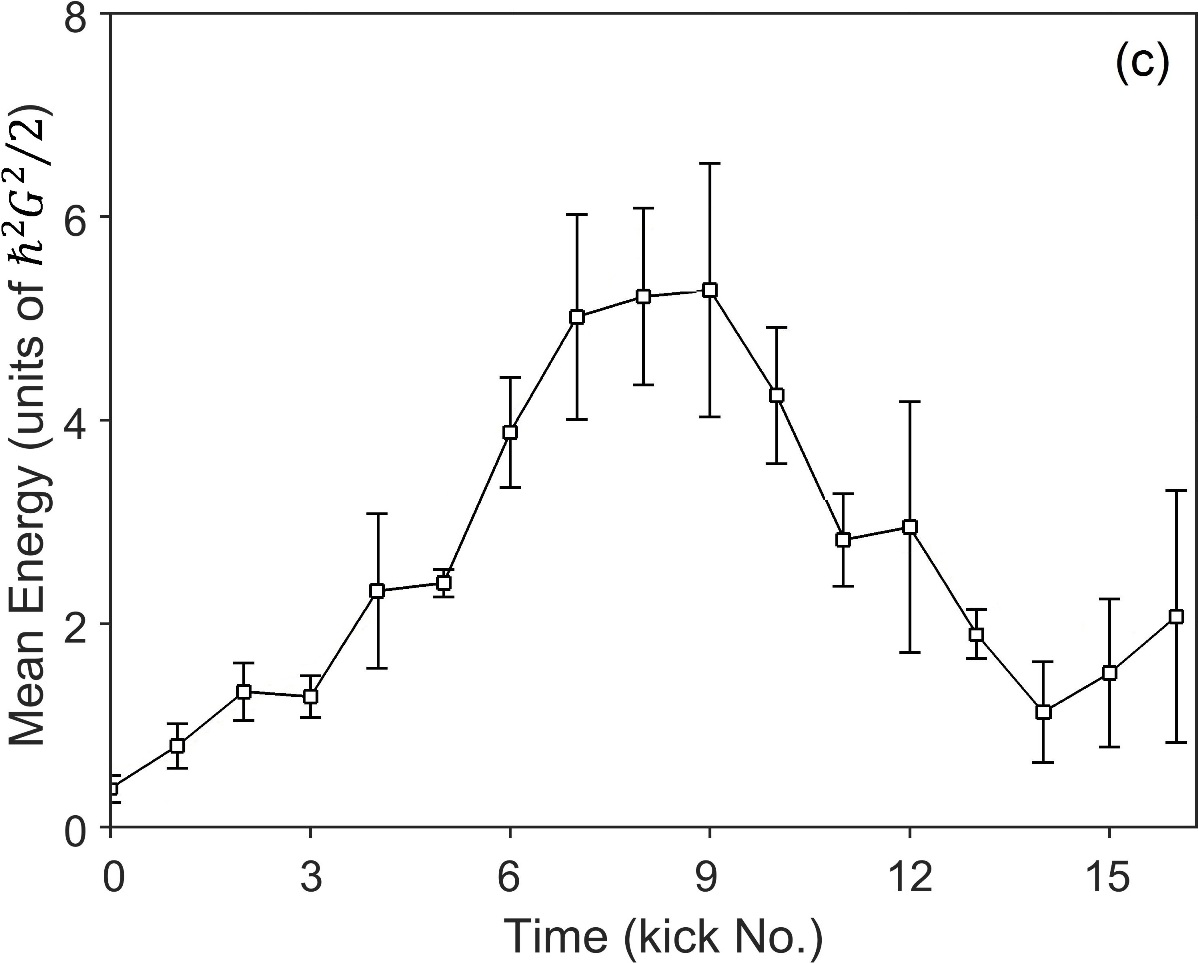}
		\caption{Experimental momentum (a) and population (b) distributions of a QW reversed at the 8th step. Panel (c) shows the variation of the mean energy corresponding to this reversal. As can be seen, the energy does not completely return to its original value. The ability to recover the initial state is experimentally limited by the contributions of various non-resonant quasimomenta that are present in the thermal cloud (and to a lesser extent the BEC).}\label{fig9}
	\end{center}
\end{figure}

Reversibility is another aspect of our momentum-based QWs. This feature is a consequence of the unitary nature of the walk and the entanglement between the internal and external degrees of freedom which can be achieved only in quantum systems. We use these properties to focus the diverging momentum currents of our QWs and retrieve the initial state of the system. Experimentally, after $j$ walk steps ${\hat{\bf{U}}}_{\rm{step}}={\hat{\bf{T}}{\hat{\bf{M}}}}$ are taken, we apply their Hermitian conjugate, ${{\hat{\bf{U}}}}^\dagger_{\rm{step}}={{\hat{\bf{M}}}}^\dagger{{\hat{\bf{T}}}}^\dagger$ for an additional $j$ steps. These conjugates can themselves be realized as $\hat{\bf{M}}^\dagger=\hat{\bf{M}}(\pi/2,\pi/2)$ and $\hat{\bf{T}}^\dagger=\hat{\bf{M}}(\pi,\pi/2)\hat{\bf{T}}\hat{\bf{M}}(\pi,-\pi/2)$ so when the intermediate ${\hat{\bf{M}}}$ matrices multiply for successive steps, the reversal steps simplify to ${{\hat{\bf{U}}}}^\dagger_{\rm{step}}=\hat{\bf{T}}\hat{\bf{M}}(\pi/2,\pi/2)$ after a $\hat{\bf{T}}\hat{\bf{M}}(\pi,-\pi/2)$ ``reflection.'' Figure~\ref{fig9} shows our experimental QW reversal with the convergence of the momentum (a) and population (b) distributions, leading to a downturn in the mean energy of the system (c). One potential application of these ``driven'' walks \cite{witthaut2010quantum,hamilton2016driven} is in quantum algorithms, e.g., searches of marked momentum states, as in \cite{sadgrove2012phase} but by adjusting the coin degree of freedom instead of the lattice. One could also make use of this reversibility in atom interferometry \cite{cronin2009optics}. The interference signal of the recombining momentum currents might be used to sense potentials affecting the phase of the system \cite{cronin2009optics,kovachy2015quantum,ahlers2016double}. 

\section{CONCLUSIONS}
\label{sec:concl.}
We have conducted a parametric study on a momentum-based QW with a $^{87}$Rb BEC. Our walk was realized using a quantum ratchet derived from an AOKR, and the coins were implemented by manipulating the two ground hyperfine levels of $^{87}$Rb with MW pulses. We elaborated distinct features of our QWs resulting from the interference and entanglement in their propagation dynamics. These walks are typically manifested by a pair of ballistically splitting peaks, differentiating them from a classical walk with a Gaussian distribution. 

Our independent access to both internal and external degrees of freedom allows us to regulate the dynamics of our QWs. We showed how the speed of our walks can be adjusted by tuning the strength of the quantum ratchet in our shift operators. We also demonstrated how the directionality of these walks can be controlled in different ways, e.g., by manipulating the superposition of the internal states (BC), changing the relative detunings from the hyperfine levels (BR), or by adjusting the phase of the gate or coin pulses accordingly. This control feature can be used for compensating probable biases in the dynamics of the quantum transport.  

In addition, by introducing a controllable noise to the phase of the system, we highlighted different stages of a quantum-to-classical transition in our walks. We also investigated the reversibility, as a direct consequence of the unitary nature of the QWs, and showed that our walks can be reversed at any step by applying the conjugates of the steps taken. This should allow the retrieval of the quantum information encoded in previous steps or even the recovery of the initial state of the system. One can also use this feature in atom interferometery \cite{ahlers2016double, kovachy2015quantum} thanks to the high sensitivity of the walk reversal to phases. Furthermore, we demonstrated how the qualitative aspects of our QWs can be affected by the global phase, introduced by the pulsed optical lattices, and fractions of the thermal cloud and various quasimomenta contributing to the BEC.

Since we implement our QWs using a BEC, our study could be extended to many-body walks \cite{preiss2015strongly,morita2016spin} by taking atom-atom interactions into account. This is interesting because a typical short-range contact interaction for BECs will act long range in momentum space \cite{alberti2017quantum, Gadway2018} offering new aspects in quantum walks of interacting particles.

In our experiment, because of the coin mixing, the external and internal degrees of freedom are effectively coupled. Such effective spin-orbital or in our context better spin-momentum coupling offers an interesting playground for testing topological properties \cite{groh2016robustness, kitagawa2010exploring, chen2014planck, chen2018observation, wang2018experimental, zhou2018floquet, cardano2017detection, Alberti2018}. For the future, it should be possible to translate the proposal from Ref. \cite{zhou2018floquet} into a feasible experiment, and to test the impact of decoherence by spontaneous emission \cite{Caspar2018b} and other noise sources such as lattice vibrations on the quantum walk and the stability of topological phases \cite{zhou2018floquet}.

Another future direction for this work is the creation of multi-dimensional QWs \citep{grafe2012biphoton,poulios2014quantum,sadgrove2012phase,chen2018observation,wang2018experimental}. The obvious way to do this would be by using a multi-dimensional optical lattice produced by laser fields propagating in different spatial dimensions. However, a tantalizing possibility with our momentum-based walks would be to employ a one-dimensional lattice with multiple spatial frequency components. In this case, each component would serve as the basis for its own momentum shift operator. As long as the spatial frequencies were not rational multiples of one another, each would produce its own distinct degree of freedom. Importantly, with a thoughtful choice of the lattices dimensions, it should still be possible to maintain quantum resonance conditions for all the spatial frequencies since the Talbot time scales inversely with the square of spatial frequency \cite{PRLGadway}. For example, two lattices with spatial frequencies that are different by a factor of $\sqrt{2}$ (an irrational number) produce two independent degrees of freedom and have Talbot times that are a factor of 2 (a rational number) apart. 


\section*{Acknowledgments}
We thank Alexander Wagner for a critical reading of the manuscript.

\bibliography{ref.bib}

\end{document}